\documentclass[article,runningheads,a4paper]{llncs}

\usepackage{amssymb}
\usepackage{graphicx}
\usepackage{listings}
\usepackage{amsmath}
\usepackage{csvsimple}
\usepackage{filecontents}
\usepackage{caption}
\usepackage{subcaption}
\usepackage[utf8]{inputenc}
\usepackage[
    backend=biber,
    style=apa,
    sorting=nyt
]{biblatex}

\usepackage{biblatex}
\usepackage{blindtext}
\usepackage{xcolor}
\usepackage{geometry}
\usepackage{hyperref}
\hypersetup{
    colorlinks=true,
    linkcolor=blue,
    citecolor=[rgb]{0,0.1,0.8},
    filecolor=blue,      
    urlcolor=blue,
    pdftitle={Overton - A bibliometric database of policy document citations},
    pdfpagemode=FullScreen,
}

\usepackage{url}
\newcommand{\keywords}[1]{\par\addvspace\baselineskip
\noindent\keywordname\enspace\ignorespaces#1}
\makeatletter
\g@addto@macro\@floatboxreset\centering
\makeatother
\begin{document} 
\mainmatter 

\title{Overton - A bibliometric database of policy document citations}

\author{Martin Szomszor\inst{1} \and Euan Adie\inst{2}}

\institute{
    Electric Data Solutions LTD, London, WC2A 2JR, United Kingdom \\
    \email{martin@electricdata.solutions}
    \and
    Overton, London, N12 2FG, United Kingdom \\
    \email{euan@overton.io}
}

\maketitle

\begin{abstract}
This paper presents an analysis of the Overton policy document database, describing the makeup of materials indexed and the nature in which they cite academic literature. We report on various aspects of the data, including growth, geographic spread, language representation, the range of policy source types included, and the availability of citation links in documents. Longitudinal analysis over established journal category schemes is used to reveal the scale and disciplinary focus of citations and determine the feasibility of developing field-normalized citation indicators. We examine how well self-reported funding outcomes collected by UK funders corresponds to data indexed in the Overton database, and if peer-review assessment of impact as measured by the UK Research Excellence Framework (REF) 2014 correlates with derived citation metrics. Our findings show that for some research topics, such as health, economics, social care and the environment, Overton contains a core set of policy documents with sufficient citation linkage to academic literature to support various citation analysis that may be informative in research evaluation, impact assessment, and policy review. The data indexed in Overton agrees with that collected via self-reporting of funding outcomes, and correlates with peer-review assessment of impact in some disciplines. 
\end{abstract}

\section{Introduction}
The premise that academic research leads to wider social, cultural, economic and environmental benefits has underpinned our investment in publicly funded research since the 1950s \parencite{bush_v_1945}. It was broadly accepted that research leads to positive outcomes \parencite{burke_1985}, but this belief was further scrutinized as technical analysis were developed to unpick the exact nature and scale of these impacts \parencite{evenson_1979}. The types of evaluation become more varied and complex as the investigators focused on specific domains \parencite{hanney_2000,van_der_meulen_2000}, taking into account the myriad ways in which knowledge is generated, exchanged, assimilated and utilized outside of academia. The general assumption holds that there is a return on investment in research through direct and indirect mechanisms \parencite{salter_2001} and the most recent literature reviews \parencite{bornmann_2013,penfield_2014,greenhalgh_2016} provide detailed perspectives on how to identify and differentiate between outputs and outcomes across a range of settings.

Research evaluation also developed to support a greater need for accountability \parencite{thomas_2020}. Initially, by peer review  \parencite{gibbons_1987}, then strategic reorientation \parencite{georghiou_1995}, and recently using more data driven approaches that incorporate bibliometric components \parencite{martin_1996,adams_2007,hicks_diana_2010,hicks_diana_2013}. Despite shortcomings in their suitability to judge research quality \parencite{moed_1985,pendlebury_2009}, citation indicators became more popular \parencite{may_1997} due to their growing availability, relatively low-cost compared with conventional peer-review, and ready application to national, regional and institutional portfolios \parencite{beis_2017}. Current evaluations programs that consider citation data include: Australia \parencite{australian_research_council_2018}, EU \parencite{reinhardt_2012}, Finland \parencite{lahtinen_2005}, Italy \parencite{abramo_2015}, New Zealand \parencite{buckle_2019}, Norway \parencite{sivertsen_2018}, Spain \parencite{jimenez-contreras_2003}, UK \parencite{ref2020_2020} and USA \parencite{national_institutes_of_health_nih_2008}.

However, growing use of bibliometric indicators also altered researcher behaviours via corrupted incentives, leading to a variety of negative outcomes \parencite{butler_2003,lopez_pineiro_2015,yucel_2018,abramo_2021} and motivating various groups to call for more nuanced and equitable research assessment, such as in the San Francisco Declaration on Research Assessment (DORA) \parencite{cagan_2013}, Metrics Tide report \parencite{wilsdon_2015}, and Leiden Manifesto \parencite{hicks_2015}. This has resulted in publishers, research organizations and funders signing-up to the aforementioned initiatives and developing their own policies to ensure metrics are deployed and used responsibly. A key aspect has been a push towards broad recognition of research contributions \parencite{morton_2015} and a more nuanced use of bibliometric indicators \parencite{adams_jonathan_2019}.

Throughout this growth and development in the use of metrics, it has become clear that standard citation indicators reflect
only the strength of influence within academia and are unable to measure impact beyond this realm \parencite{ravenscroft_2017,moed_2005}. This has led to the exploration of adjacent data sources to provide signals of the wider impact of research, which have been collectively named altmetrics \parencite{priem_j_2010}. This term refers to a range of potential data sources that could potentially reveal educational impact \parencite{mas-bleda_2018,kousha_2008}, knowledge transfer \parencite{kousha_2017}, commercial use \parencite{orduna-malea_2017}, public engagement \parencite{shema_2015}, policy influence \parencite{tattersall_2018}, and more. With access to a broader range of indicators, it may be possible to address some contemporary research evaluation issues by increasing the scope of how research is measured and allow the full range of research outcomes to be attributed to researchers.

In the area of policy influence, the research underpinning clinical guidelines, economic policy, environmental protocols, etc. is a significant topic of interest. Analysis of the REF2014 Impact Case Study data \parencite{grant_jonathan_2015} showed that 20\% of case studies were associated with the topic \textit{Informing government policy}, and 17\% were associated with \textit{Parliamentary scrutiny}, most frequently in Panel C (social sciences). In many cases, evidence cited in case studies included citations to the research from national and international policy organizations. In Unit of Assessment 1 (clinical medicine), 41\% of case studies were allocated to the topic \textit{Clinical guidance} indicating some use of the academic research in policy setting.

Since 2019, a large database of policy documents and their citations to academic literature has been developed by Overton (see \href{https://www.overton.io/}{overton.io}). It currently (as of December 2021) indexes publications from more than 30,000 national and international sources including governments, think tanks, intergovernmental organizations (IGOs) and charities. The focus of this paper is to evaluate Overton as a potential bibliometric data source using a series of analysis that investigate the makeup of documents indexed (e.g. by geography, language and year of publication), the network of citations (e.g. volume, distribution, time-lag), how well data correlate with other impact logging processes (e.g. as reported to funders), and if derived citation metrics correlate with peer-review assessment. In doing so, it is our hope to understand more about the potential uses of policy citation data by highlighting which disciplines are most frequently cited and if citation volumes are sufficient to support the development of citation indicators.

The paper is structured as follows: Section \ref{sec:relatedwork} summarizes related work, Section \ref{sec:methodology} presents the methodology for each experiment and outlines the datasets used. Section \ref{sec:results} presents the results of each analysis before discussion in Section \ref{sec:discussion}.

\section{Related Work} \label{sec:relatedwork}
The traditional bibliometric databases, namely the Web of Science (Clarivate), Scopus (Elsevier), Dimensions (Digital Science), Microsoft Academic (Microsoft), and Google Scholar (Google), have been extensively evaluated \parencite{visser_2021,aksnes_2019,chadegani_2013,falagas_2008,harzing_2016}, particularly in terms of cited references \parencite{martin-martin_2021}, subject coverage \parencite{martin-martin_2018}, comparability of citation metrics \parencite{thelwall_2018}, journal coverage \parencite{mongeon_2016,singh_2021}, classification systems \parencite{wang_2016}, accuracy of reference linking \parencite{alcaraz_2012,olensky_2016}, duplication \parencite{valderrama-zurian_2015}, suitability for application with national and institutional aggregations \parencite{guerrero-bote_2021}, language coverage \parencite{vera-baceta_2019}, regional bias \parencite{tennant_2020,rafols_2020}, and predatory publishing \parencite{bjork_2020,demir_2020}. The notion of best data source is partly subjective (i.e. depending on personal preference), but  also depends on the type of use (e.g. search and discovery versus bibliometric analysis), discipline, regional focus, time period in question, and can be influenced by the availability of metadata and links to adjacent datasets (e.g. patents, grants, clinical trials, etc), depending on task.

Much like the preference for bibliographic data source, the choice of citation impact indicator \parencite{waltman_2016} is highly debatable. It is generally accepted that citations should be normalized by year of publication, discipline, and document type, although whether the calculation should be based on the average of ratios \parencite{waltman_2011,opthof_2010} or ratio of averages \parencite{moed_2010,vinkler_2012} is contentious \parencite{lariviere_2011}, as is the selection of counting methodology \parencite{waltman_2015,potter_2020}. Suitable sample size is key to providing robust outcomes \parencite{rogers_2020}, and any choices made with respect to category scheme used and indicator choice should influence interpretation of results \parencite{szomszor_2021}.

The potential for use of altmetric indicators was initially focused on the prediction of traditional citations \parencite{thelwall_2013} and possible correlation with existing indicators \parencite{zahedi_2014,costas_2015}. It was suggested that \textit{``little knowledge is gained from these studies''} \parencite{bornmann_2014a} and that the biggest potential for altmetrics was toward measurements of broader societal impact \parencite{bornmann_2015}. At this point, the coverage of altmetrics was limited to social media attention (e.g. Twitter and Facebook mentions), usage metrics (e.g. website downloads, Mendeley readers), and online news citations (both traditional and blogs). Comparisons with peer-review assessment \parencite{bornmann_2018} revealed that Mendeley readership was most strongly associated of these with high quality research, but still much less than conventional citation indicators. More recent analysis \parencite{bornmann_2019} have incorporated other altmetric indicators showing Wikipedia and policy document citations to have the highest correlation with REF Impact Case study scores out of the available indicators. \parencite{bornmann_2016} concludes \textit{``Policy documents are one of the few altmetrics sources which can be used for the target-oriented impact measurement''}. The only research to date that uses the Overton policy document database 
\parencite{pinheiro_2021} finds that cross-disciplinary research will increase the policy relevance of research outcomes.

Prior work investigating the translation of research through citations in clinical guidelines \parencite{grant_2000} have utilized specific data sources and other manually curated data sets \parencite{kryl_2012,newson_2018} to show their value in evaluating research outcomes. Databases of clinical practice guidelines have emerged \parencite{eriksson_2020} to support this specific line of enquiry, and recent work \parencite{daraio_2020,pallari_2021,guthrie_2019} utilizes this information to uncover national trends and highlight relative differences in the evidence base used.

\section{Methodology} \label{sec:methodology}
The Overton database is the primary source of data for this study. It is created by web-crawling publicly accessible documents published by a curated list of over 30,000 organizations including governments, intergovernmental organizations, think tanks, and charities. Each document is processed to extract bibliographic information (title, authors, publication date, etc) along with a list of cited references, including those to academic literature as well as other policy documents. A policy document itself may be composed of multiple items, referred to herein as PDFs since they are the majority format type, such as clinical guidelines (which contain separate documents with recommendations and evidence bases) or when language translations exist. The types of documents vary in nature and include reports, white papers, clinical guidelines, parliamentary transcripts, legal documents, and more.

Overton classifies publication sources using a broad taxonomy that is further sub-divided by type. Top level source types are: government, igo, think tank, and other. Sub-types include bank, court, healthcare agency, research centre, and legislative. Each publication source is assigned a geographic location, including country and region (e.g. state or devolved territory). Some sources are classified as IGO (i.e. global reach), or EU (European Union).

For this study, 4,504,896 policy documents (made up of 4,854,919 individual PDFs) citing 3,579,710 unique articles (DOIs) were used. To integrate this data with other sources, all records were converted into Resource Description Framework (RDF) \parencite{bizer_2018}, a semantic web metadata model, and loaded into a the graph database GraphDB\texttrademark. The following additional data sources were used:
\begin{itemize}
    \item \textbf{Crossref} - metadata for all DOIs were extracted from Crossref records providing titles, source names (i.e. journal), collection identifiers (ISSNs and ISBNs), and publication dates.
    \item \textbf{Scopus journal categories} - As determined by linking ISSNs to Crossref records, each journal is associated with up to 13 All Science Journal Classification (ASJC) categories for journals, organized in a hierarchy under areas and disciplines. ($n=19,555$).
    Source: \href{https://www.scopus.com/sources?dgcid=RN_AG_Sourced_300005499}{scopus.com}
    \item \textbf{REF2014 Case Studies} - all publicly available case studies submitted to REF2014 and the associated DOIs mentioned in the references section. A total of 6,637 Case Studies were included, linking to 24,945 unique DOIs. Source: \href{https://impact.ref.ac.uk/casestudies/}{impact.ref.ac.uk}
    \item \textbf{REF2014 Results} - the final distribution of scores awarded in REF2014. For each Institution and UoA, scores for Outputs and Case Studies were loaded, expressed as the percentage of outputs in categories 4* (world-leading), 3* (internationally excellent), 2* (internationally recognized), and 1* (nationally recognized). Source: \href{https://results.ref.ac.uk/(S(4sccogdjgp4n33h010w4pt3e))/DownloadResults}{results.ref.ac.uk}
    \item \textbf{Gateway to Research (GTR)} - all funded projects from UKRI Research Councils ($n=123,399$), their associated publications ($n=1,015,664$), and outcomes categorized as policy outcome ($n=39,406$). 
    Source: \href{https://gtr.ukri.org/resources/GtR-2-API-v1.7.5.pdf}{gtr.ukri.org}
\end{itemize}

This combination of information allows us to investigate a range of questions that will inform the potential viability of Overton as a bibliometric data source:
\begin{enumerate}
\item \textit{What is the makeup of the database in terms of sources indexed by geography, language, type and year of publication?} - this analysis will determine by year of publication, the count of policy documents and PDFs indexed according to source type, region, country and language. This will reveal potential biases in coverage that would inform suitability for certain types of analysis. Overton does contain locally relevant policy sources, such as regional government publications, but not for all geographies.
% figs 1-3

\item \textit{How many scholarly references are extracted and over what time period?} - will measure the total number of references to DOIs extracted according to policy publication year and source type, and show the count of citations received to DOIs by their publication year according to broad research area. It is important to know how many citations to research articles are tracked because the volume will inform their suitability for citation-based indicator development.
% figs 4-5, table 1

\item \textit{How long does it take research articles to accumulate policy citations and how does this vary across disciplines?} - will provide details on how long DOIs take to accumulate citations, both in absolute volume per year, and cumulatively. Research areas and disciplines will be analysed separately to illustrate any differences and to highlight domains in which citation analysis may be fruitful.
% figs 6-7

\item \textit{What is the time-lag between the publication of scholarly works and their citation within policy literature and how does this vary between disciplines?} - will show the distribution from the perspective of citing policy document (i.e. how old are cited references), and from cited DOI (i.e. when are citations to research articles received). A sample of policy sources for healthcare agencies and governmental banks are also benchmarked to illustrate feasible comparisons. The range and timeliness of evidence used is an important consideration in policy evaluation and may be possible using the Overton database.
%figs 8-9

\item \textit{What statistical distribution best models policy citation counts to research articles?} - will test the fit of various distributions (e.g. powerlaw, lognormal, exponential) to empirical data using conventional probably distribution plots. Analysis by research discipline and subject will be used to inform potential field-based normalization techniques (i.e. appropriate level of granularity).
%figs 10-11

\item \textit{How feasible is field-based citation normalization?} - will determine if a minimum sample size can be created for each subject category and year for DOIs published between 2000-20. This analysis will highlight subjects that may be suitable for citation metrics and those where insufficient data are available to make robust benchmarks.
% figs 12 + table 2

\item \textit{Do the citations tracked in the policy literature correlate with policy influence outcomes attributed to funded grants?} - will test the correlation between policy influence outcomes reported against funded grants (submitted via the ResearchFish platform to UKRI), and the number of Overton policy citations from DOIs specified as outputs of these projects. Correlations will also be calculated for each subject according to the GTR classification.
% fig 13 + table 4

\item \textit{Does the amount of policy citation correlate with peer-review assessment scores as reported in the UK REF2014 impact case study data?} - will test size-independent correlation \parencite{traag_2019} between normalized policy citation metrics (percentiles) and peer-review assessment (according to 4* rating). Percentiles are calculated based on year of publication and Scopus ASJC subject categories.
% fig 14

\end{enumerate}
To analyse data by research subjects, disciplines, and areas, we utilize the Scopus ASJC journal subject mapping. This is the preferred categorical system for this analysis because it is matched to the highest number of journals in the dataset (compared to Web of Science journal categories or the ScienceMetrix journal classification), and offers three-levels of aggregation ($areas \rightarrow disciplines \rightarrow subjects$).

\section{Results} \label{sec:results} 

\subsection{What is the makeup of the database in terms of sources indexed (by geography, language, type and year of publication)?}
\begin{figure}[!t]
     \centering
     \begin{subfigure}[b]{.49\textwidth}
         \centering
         \includegraphics[width=\textwidth]{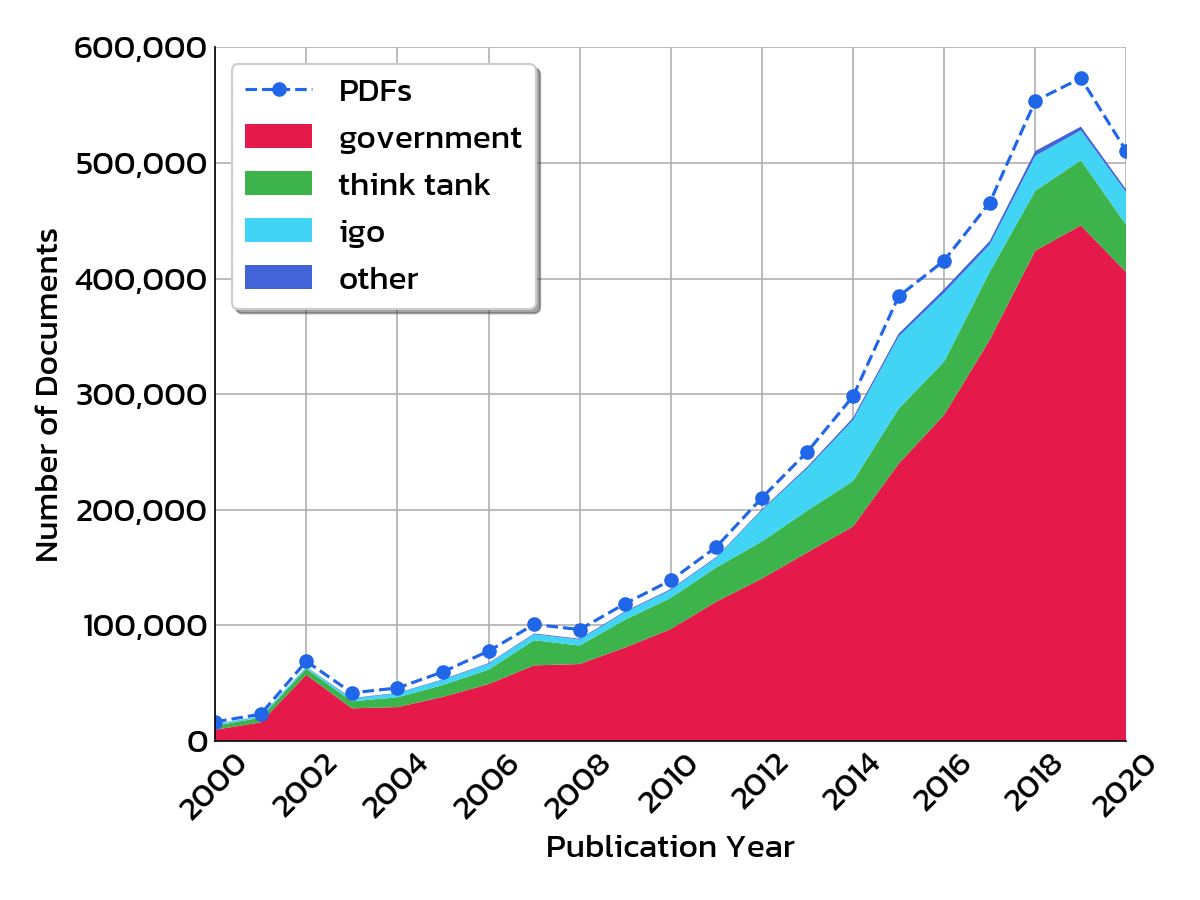}
         \caption{Document count by source type}
         \label{fig:1a}
     \end{subfigure}
     \hfill
     \begin{subfigure}[b]{.49\textwidth}
         \centering
         \includegraphics[width=\textwidth]{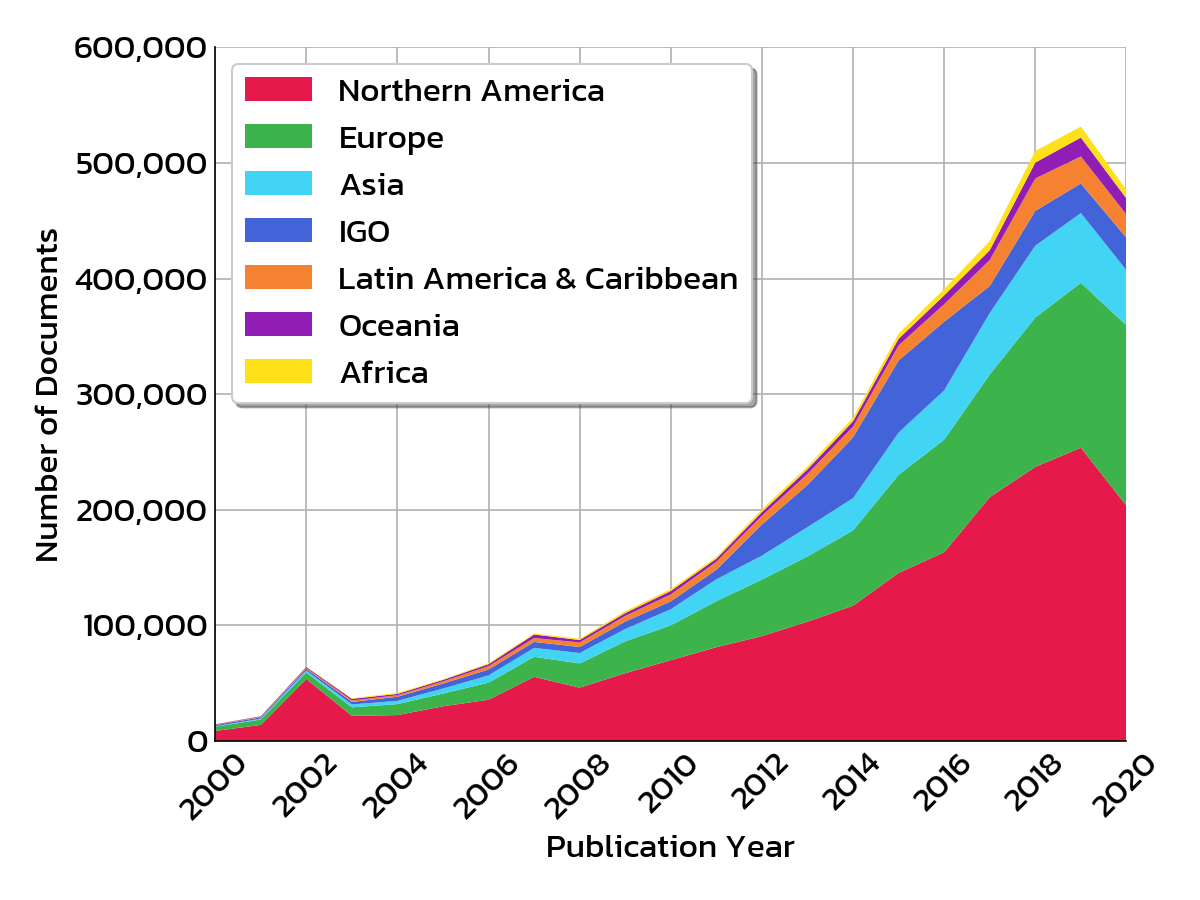}
         \caption{Document count by source region}
         \label{fig:1b}
     \end{subfigure}
     \vfill
     \begin{subfigure}[b]{.49\textwidth}
         \centering
         \includegraphics[width=\textwidth]{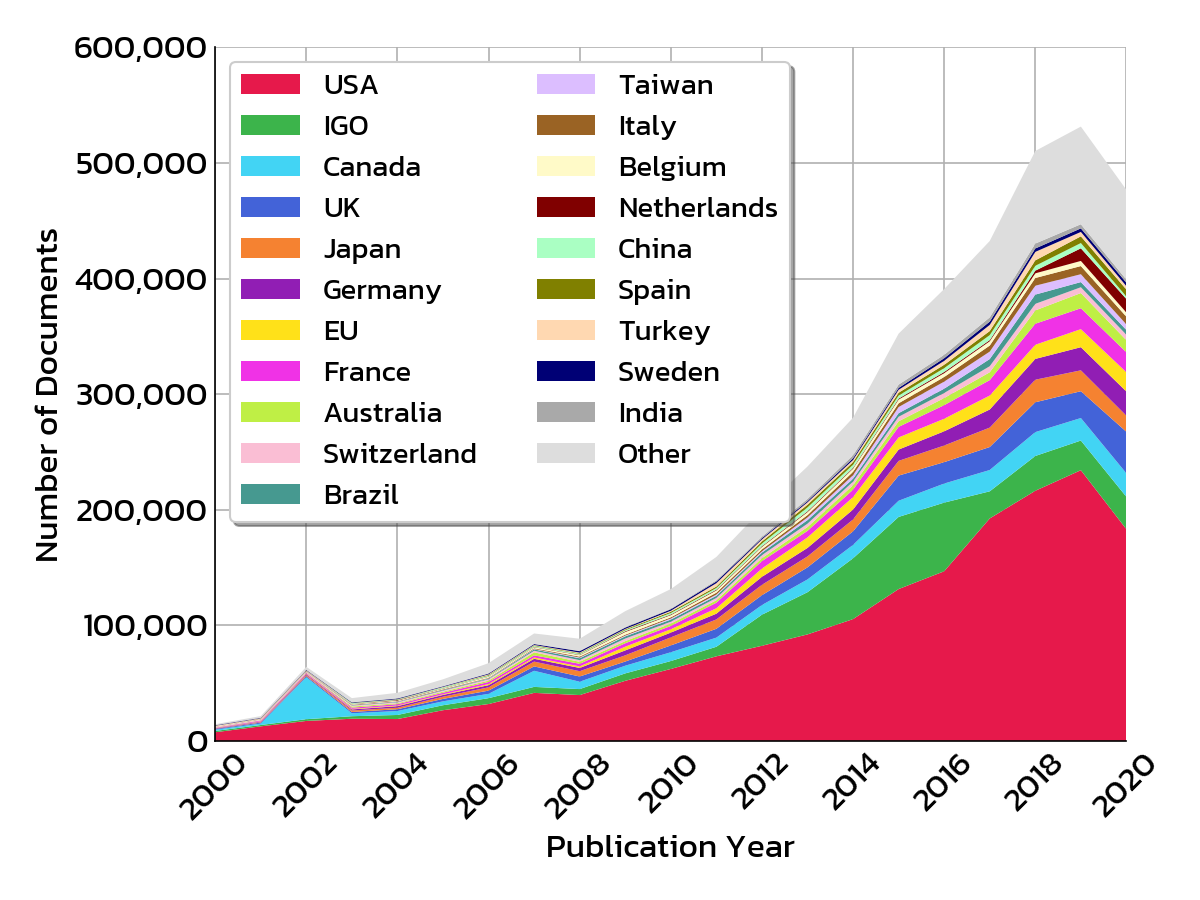}
         \caption{Document count by source country}
         \label{fig:1c}
     \end{subfigure}
     \hfill
     \begin{subfigure}[b]{.49\textwidth}
         \centering
         \includegraphics[width=\textwidth]{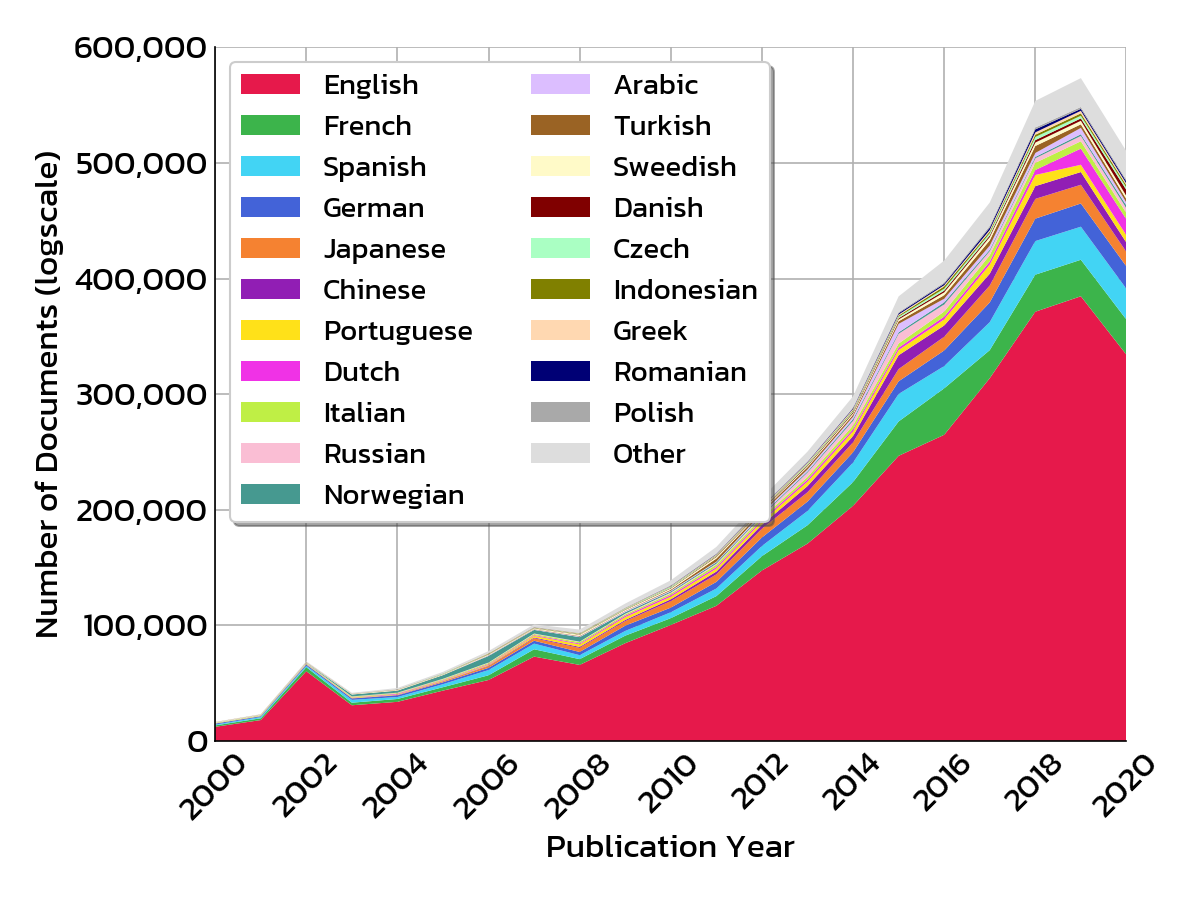}
         \caption{Document count by language}
         \label{fig:1d}
     \end{subfigure}     
     
    \caption{Time-series of Overton policy document count}
    \label{fig:1}     
\end{figure}

% figs 1-3
The growth of documents indexed in Overton is depicted in Figure \ref{fig:1}. Four plots are included showing \ref{fig:1a} - the number of documents according to publication source type (government, think tank, igo and other), \ref{fig:1b} - the number of documents indexed according to publication source region, \ref{fig:1c} - by publication source country (top 20), and \ref{fig:1d} publication language (top 20). As mentioned earlier, a policy document may contain multiple PDFs, typically language translations or different parts of a larger report or set of guidelines. The total number of PDFs indexed is shown with a dotted line in Figure \ref{fig:1a} which also corresponds to the total in Figure \ref{fig:1d} since PDFs are associated with languages rather than the policy document container (i.e. a single policy document may exist is multiple languages as different PDFs). It should be noted that while there is a significant growth in the total number of documents indexed, this doesn't necessarily correlate to a growth in the publication of policy documents overall - it only reflects how many resources are currently discoverable on the web. In this sense, our analysis shows that the availability of data is improving.

\begin{figure}[!t]
\centering
\includegraphics[width=\textwidth]{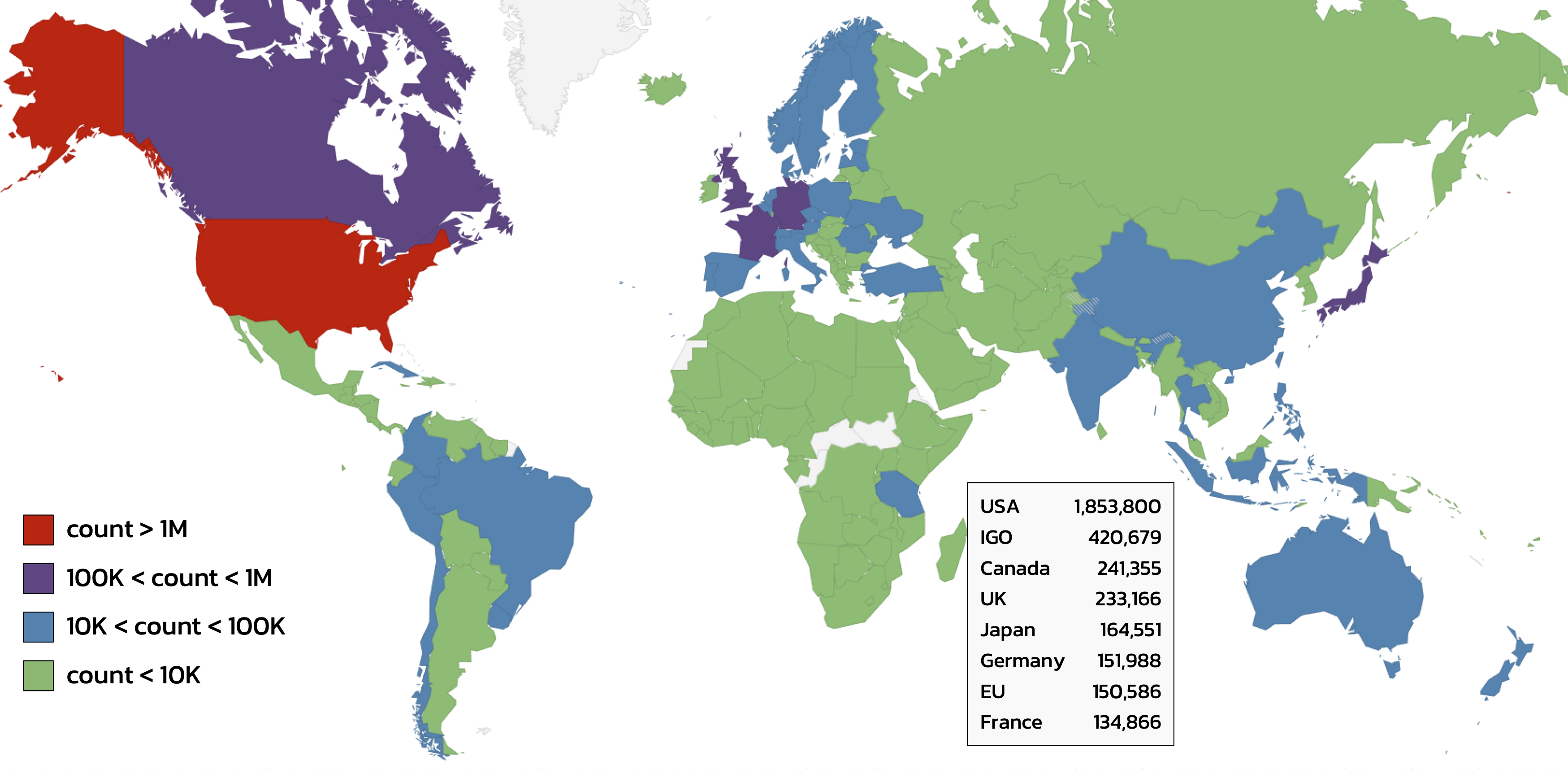}
\caption{Map showing the volume of policy documents indexed by country}
\label{fig:2}
\end{figure}

To illustrate global coverage, we also supply a map in Figure \ref{fig:2}. The map includes an insert showing the number of documents indexed for the top eight regions. Due to the scale difference between the large number of document indexed from the USA compared with other countries, four colour bins are used rather than a straightforward linear gradient.

\begin{figure}[!t]
\centering
\includegraphics[width=\textwidth]{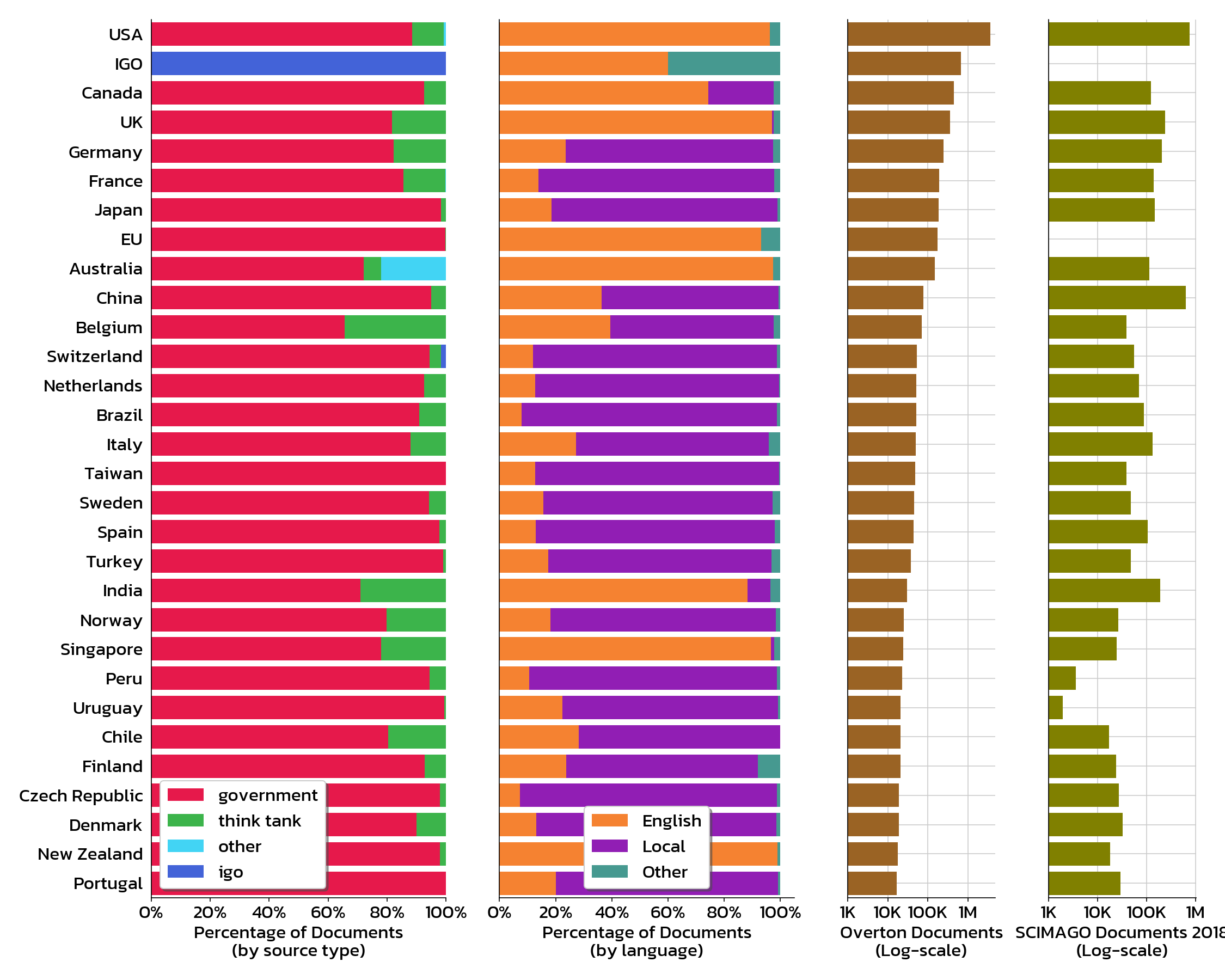}
\caption{Make up of policy documents by country}
\label{fig:3}
\end{figure}

Clearly, Overton is dominated by policy documents published by sources in the USA, but also includes significant coverage for Canada, the UK, Japan, Germany, France and Australia with the majority of content originating from governmental sources. The IGO grouping (including organizations such as the WHO, UNESCO, World Bank, and United Nations), and European Union also makes up a sizable portion of the database. In terms of the makeup of sources and languages, Figure \ref{fig:3} is included to show the percentage makeup of documents from the top 30 regions according to source type (left) and language (middle-left). For language, three values are shown: those in English, those in a local language, and those in other languages. For the regions IGO and EU, no local languages are specified. For reference, the total policy document count for each is shown (middle-right, log-scale), along with the 2018 count of articles attributed to the country in the SCImago journal ranking.

The balance of source types in each country does vary, with some regions almost entirely represented by governmental sources, such as Japan, Taiwan, Turkey and Uruguay. The unusually high percentage of documents from Australian sources categorized as other is due to articles indexed from the Analysis \& Policy Observatory (also known as APO). Another large aggregator, PubMed Central, is also indexed by Overton (for practice and clinical guidelines), but is attributed to the country USA and hence, only appears as small fraction of their output which is very large overall. 

In terms of language balance, many countries have a significant proportion of content in local languages - more than 80\% for France, Japan, Switzerland, Netherlands, Brazil, Taiwan, Sweden, Spain, Norway, Peru, Czech Republic, and Denmark. Those that do not are either English speaking (USA, UK, Australia, New Zealand) or have strong colonial ties (India and Singapore). 

The comparison of Overton content to SCIMago article count is included to show possible over and under representation. For example, China produces the second largest number of academic articles (after the USA), but is only the 8\textsuperscript{th} most frequently indexed country (excluding IGO and EU) in Overton. In contrast, Peru and Uruguay produce a much lower number of research articles than Brazil and Chile, but a similar amount of content is indexed in Overton.

\subsection{How many scholarly references are extracted and over what time period?}
\begin{figure}[!t]
     \centering
     \begin{subfigure}[b]{.49\textwidth}
         \centering
         \includegraphics[width=\textwidth]{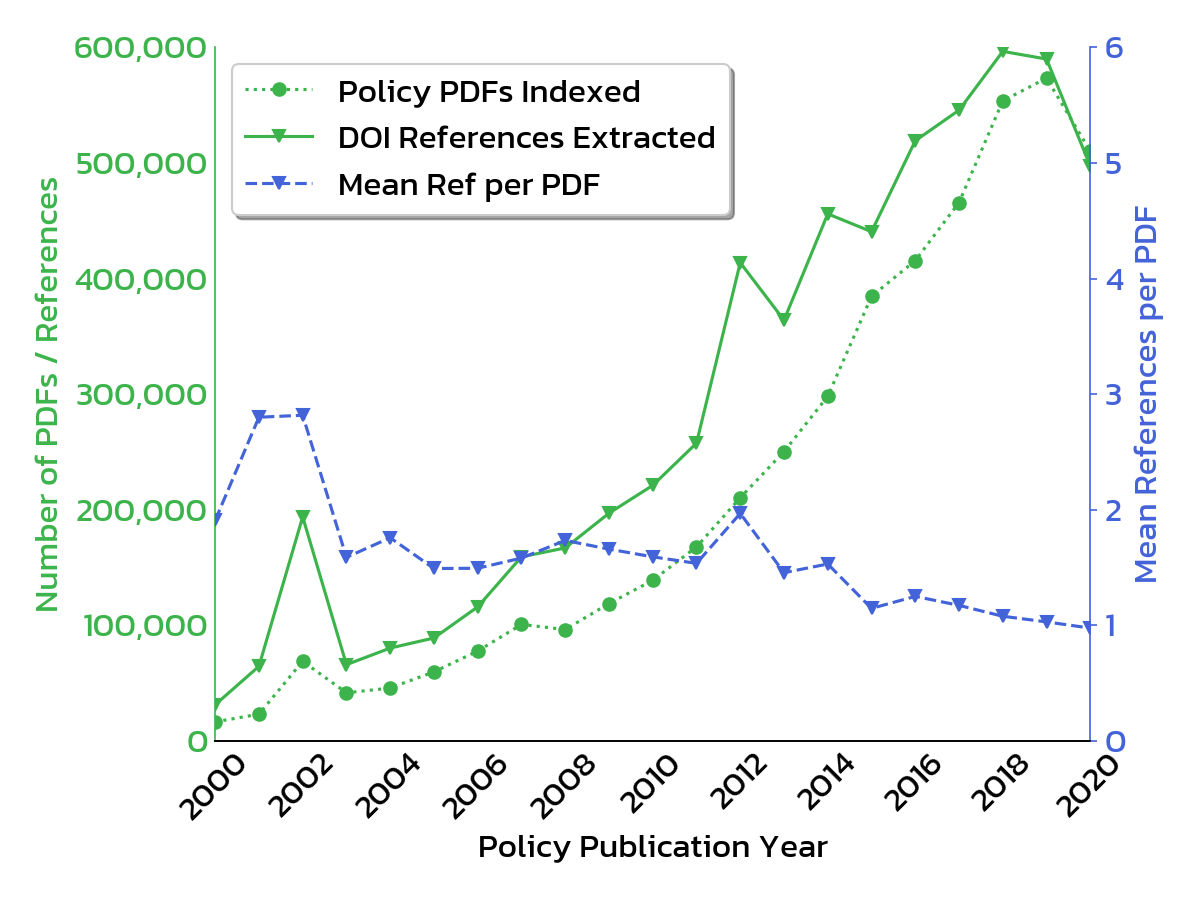}
         \caption{Global count of references extracted from policy PDFs (left) and running average (right)}
         \label{fig:4a}
     \end{subfigure}
     \hfill
     \begin{subfigure}[b]{.49\textwidth}
         \centering
         \includegraphics[width=\textwidth]{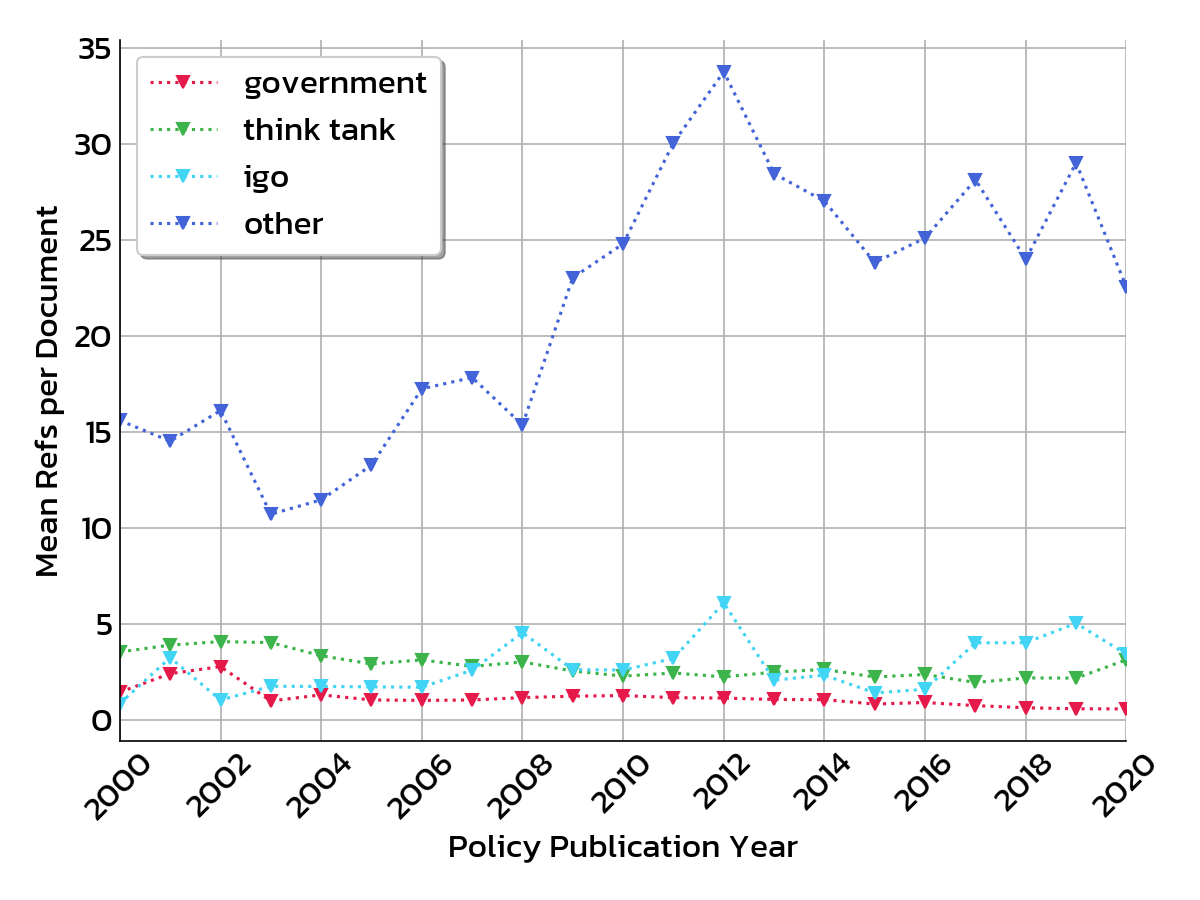}
         \caption{Average number of references extracted by source type}
         \label{fig:4b}
     \end{subfigure}
    \caption{Time-series of Policy Document references extracted}
    \label{fig:4}     
\end{figure}

% figs 4-5, table 1
For each PDF indexed by Overton, references to research literature are identified and extracted. The number of PDFs indexed and the corresponding number of scholarly references extracted are shown for each year in the period 2000-2020 in Figure \ref{fig:4a}. Only references to DOIs are included in this analysis - $2,027,440$ references to other policy documents are excluded. The left axis (green) shows the totals and the right axis (blue) shows the average number of references per PDF. This data is also broken down by publication source type in Figure \ref{fig:4b} where the average (mean) is shown for each through the period 2000-2020. The type `other' includes articles from PubMed Central which would account for the relatively high rate of reference extraction for that source type compared to others, albeit for a small fraction of the database (about 1\% of PDFs). 

Data are also summarized in Table \ref{table:1} where each row corresponds to a set of policy PDFs that contain a minimum number of scholarly references. For example, row $>=10$ counts all PDFs that have 10 or more references to scholarly articles. There are $214,082$ of these ($4.4\%$ of the corpus) that account for a total of $8,633,884$ reference links, or $89\%$ of references overall. This data indicates that although there are many policy documents that have no references, a core set of documents (approximately 200,000) may contain a sufficient number of references to build useful citation indicators. It is also possible that the documents that have no references may be linked to other entities in Overton, such as researchers, institutions and topics of interest, providing other analytical value. 
{\setlength{\tabcolsep}{0.3em}
\begin{table}[!h]
    \centering
    \begin{tabular}{ |l||r|r|r|r|}
    \hline    
     Refs. Count & PDFs & \% PDFs & Total Refs. & \% Refs. \\ 
    \hline
$\geq0$    & 4,854,919 &     100.00 & 9,747,436 &     100.00 \\
$\geq1$    &   570,830 &      11.76 & 9,747,436 &     100.00 \\
$\geq5$    &   305,637 &       6.30 & 9,248,600 &      94.88 \\
$\geq10$   &   214,082 &       4.41 & 8,633,884 &      88.58 \\
$\geq50$   &    38,235 &       0.79 & 4,772,402 &      48.96 \\
$\geq100$  &    14,162 &       0.29 & 3,139,856 &      32.21 \\
$\geq500$  &       794 &       0.02 &   725,307 &       7.44 \\
$\geq1000$ &       181 &       0.00 &   312,596 &       3.21 \\
    \hline
    \end{tabular}
    \caption{Count and percentage of scholarly references made from Policy PDFs by reference count category}
    \label{table:1}
\end{table}}

\begin{figure}[!t]
     \centering
     \begin{subfigure}[b]{.49\textwidth}
         \centering
         \includegraphics[width=\textwidth]{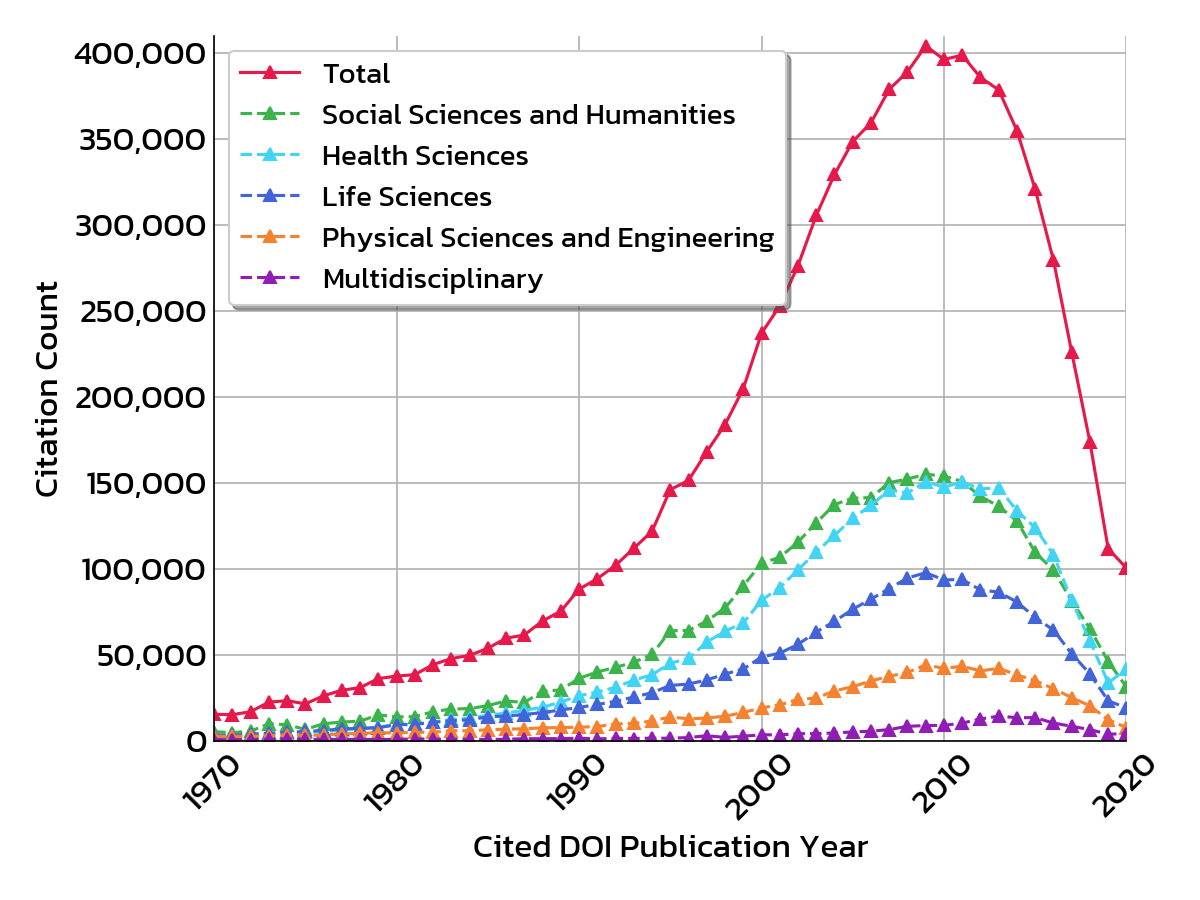}
         \caption{Number of Citations to DOIs - total and by research area}
         \label{fig:5a}
     \end{subfigure}
     \hfill     
     \begin{subfigure}[b]{.49\textwidth}
         \centering
         \includegraphics[width=\textwidth]{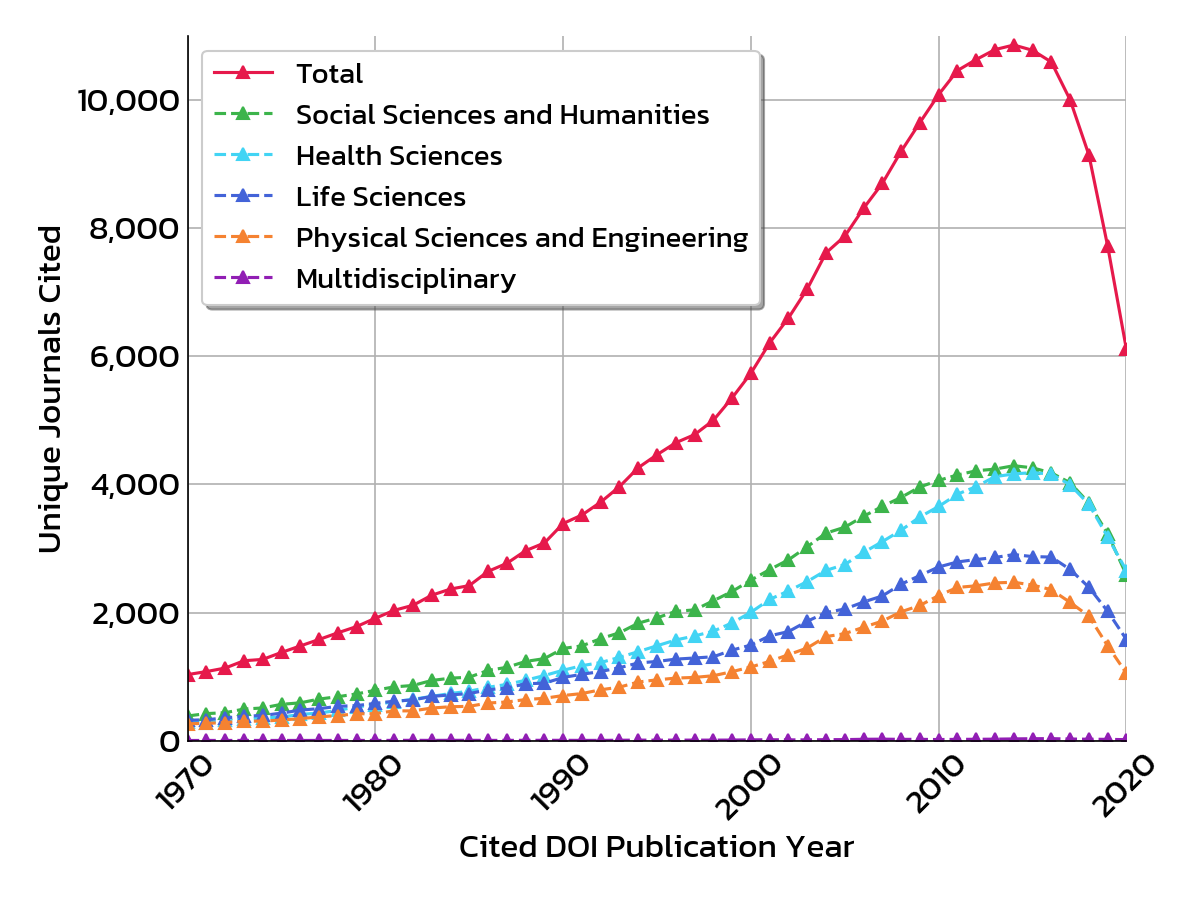}
         \caption{Number of unique Journals cited - total and by research area}
         \label{fig:5b}
     \end{subfigure}

    \caption{Time-series of citation counts to DOIs and unique journals.}
    \label{fig:5}     
\end{figure}

Perhaps of more interest from the perspective of building citation indicators, Figure \ref{fig:5a} presents the number of citations received by DOIs according to their year of publication, dating back to 1970. The database total is shown in red, along with the corresponding totals for main research areas (as defined by ASJC). The data shows that since 2000, publications have been cited in each year at least $200,000$ times, with a maximum of $404,271$ in 2009. We also use the same data to plot Figure \ref{fig:5b} which shows the number of unique journals receiving citations in each year. The total maximum of around 10,000 corresponds well with the core set of global journals, for example in the Web of Science flagship collection or core publication list in the Leiden Ranking \parencite{leiden_rank_2021}.

\subsection{How long does it take research articles to accumulate policy citations and how does this vary across disciplines?}
\begin{figure}[!b]
     \centering
     \begin{subfigure}[b]{.49\textwidth}
         \centering
         \includegraphics[width=\textwidth]{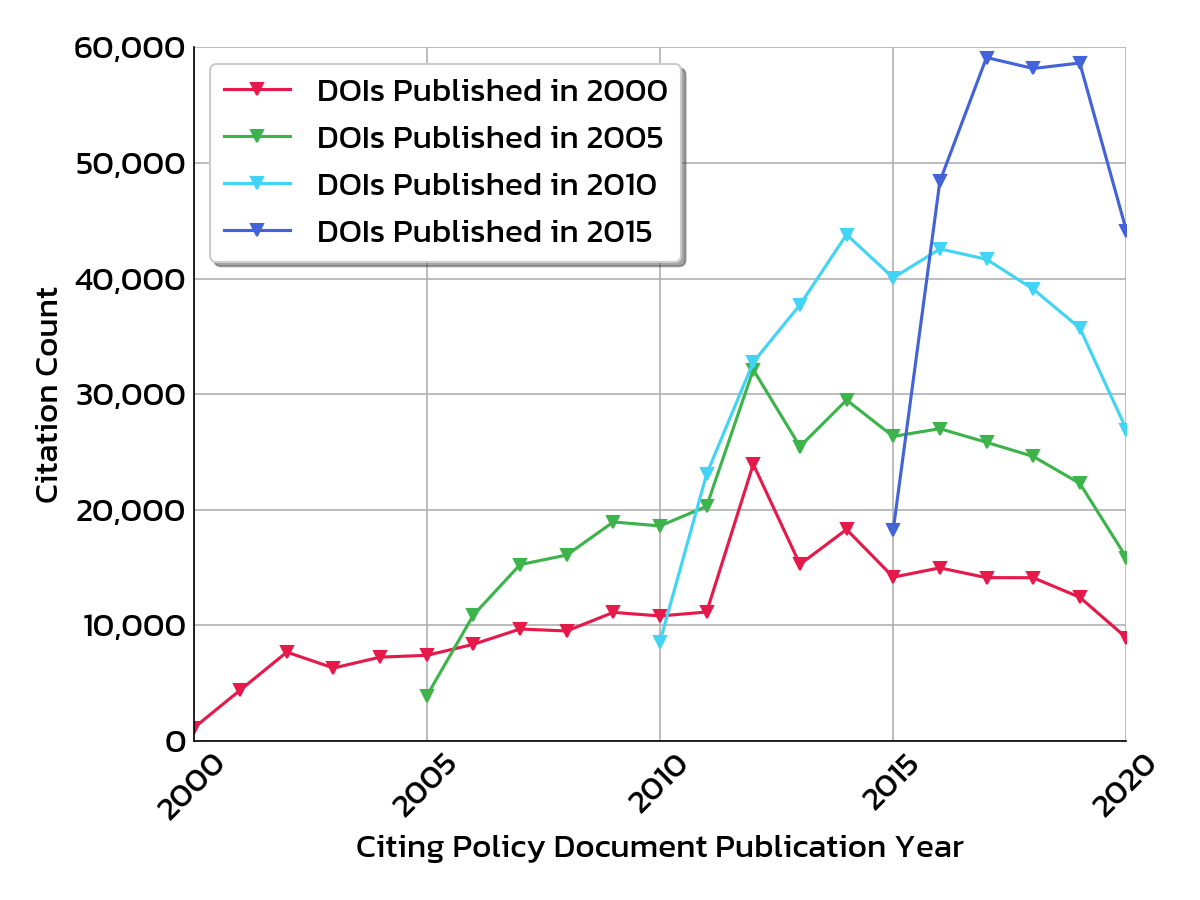}
         \caption{Number of citations received by DOIs in each subsequent year}
         \label{fig:6a}
     \end{subfigure}
     \hfill     
     \begin{subfigure}[b]{.49\textwidth}
         \centering
         \includegraphics[width=\textwidth]{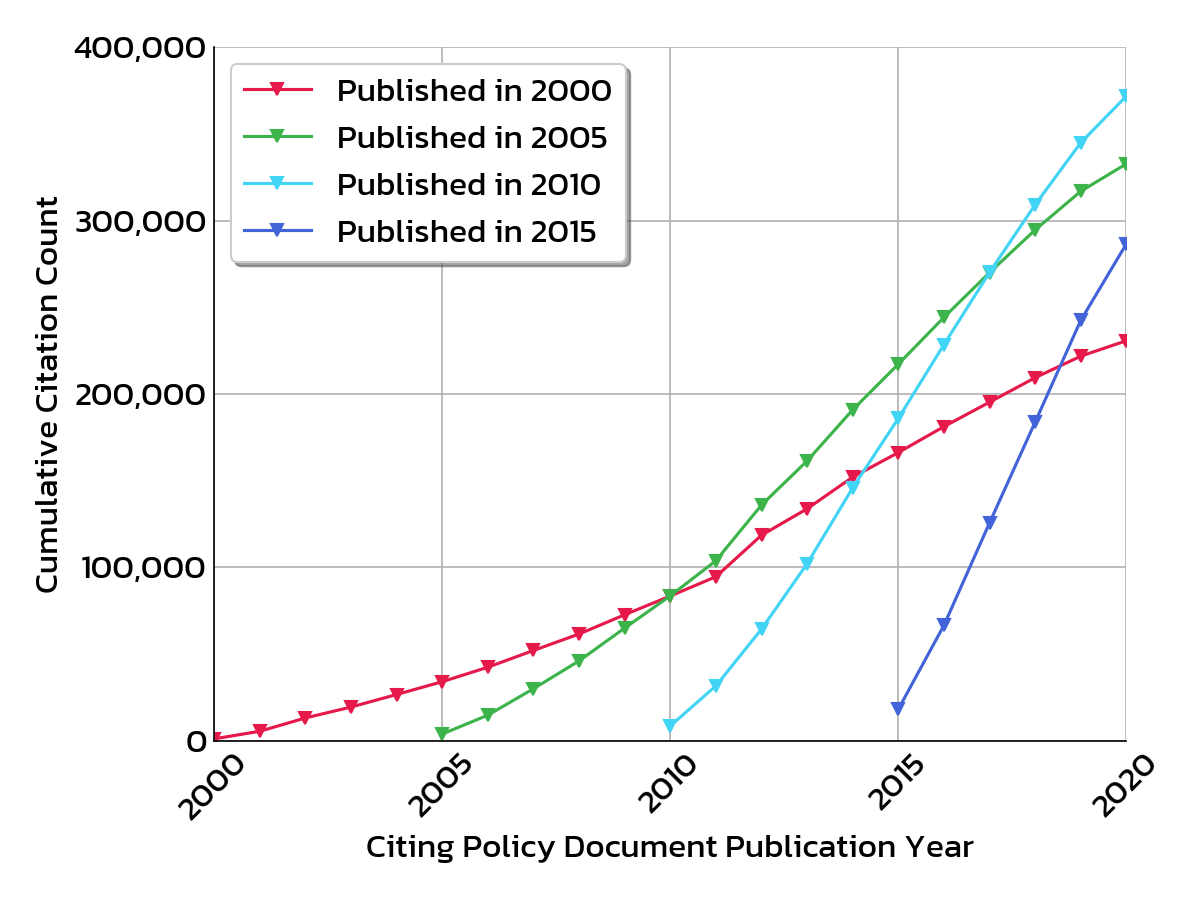}
         \caption{Cumulative count of citation received by DOIs in each subsequent year}
         \label{fig:6b}
     \end{subfigure}

    \caption{Time-series (total and cumulative) of citations received to DOIs published in 2000, 2005, 2010, and 2020}
    \label{fig:6}     
\end{figure}

% figs 6-7
To appreciate the dynamics of how research articles accumulate citations from policy literature, we plot the number of citations received in years following original publication for DOIs published in 2000, 2005, 2010, and 2015. In Figure \ref{fig:6a}, the total number of citations received in each year is plotted, and in Figure \ref{fig:6b}, the cumulative total is displayed. These data indicate that the citation lifetime for DOIs is not even across years - older publications have received fewer citations overall and over a longer time period than those published more recently. Articles published in 2005 peaked seven years after publication, those published in 2010 peaked after four years, and those published in 2015 after only two years. Further investigation is necessary to understand these differences, but it might be accounted for by the way the database is growing - an increasing number of documents indexed year-on-year could manifest as a recency bias.

\begin{figure}[!t]
     \centering
     \begin{subfigure}[b]{.49\textwidth}
         \centering
         \includegraphics[width=\textwidth]{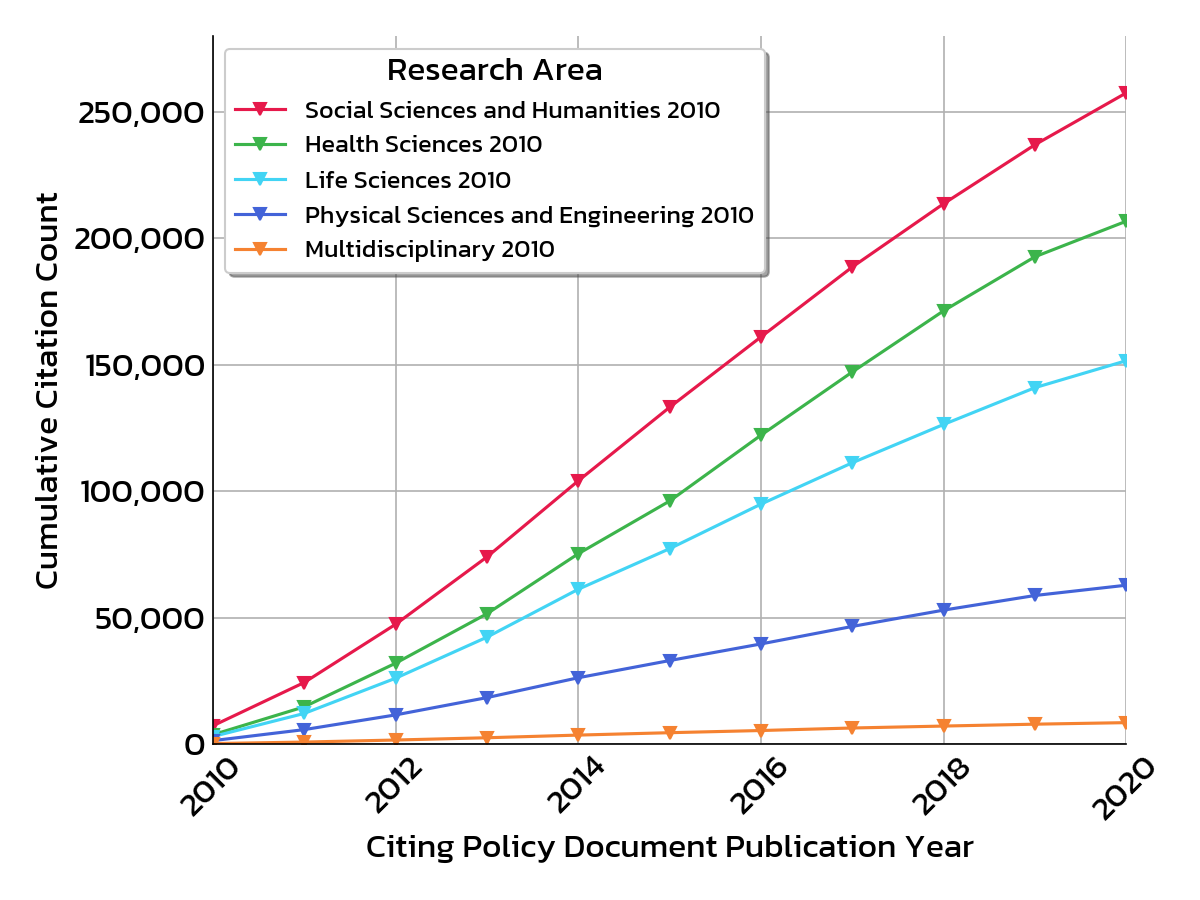}
         \caption{Number of Citations to DOIs published in 2010 by research area}
         \label{fig:7a}
     \end{subfigure}
     \hfill     
     \begin{subfigure}[b]{.49\textwidth}
         \centering
         \includegraphics[width=\textwidth]{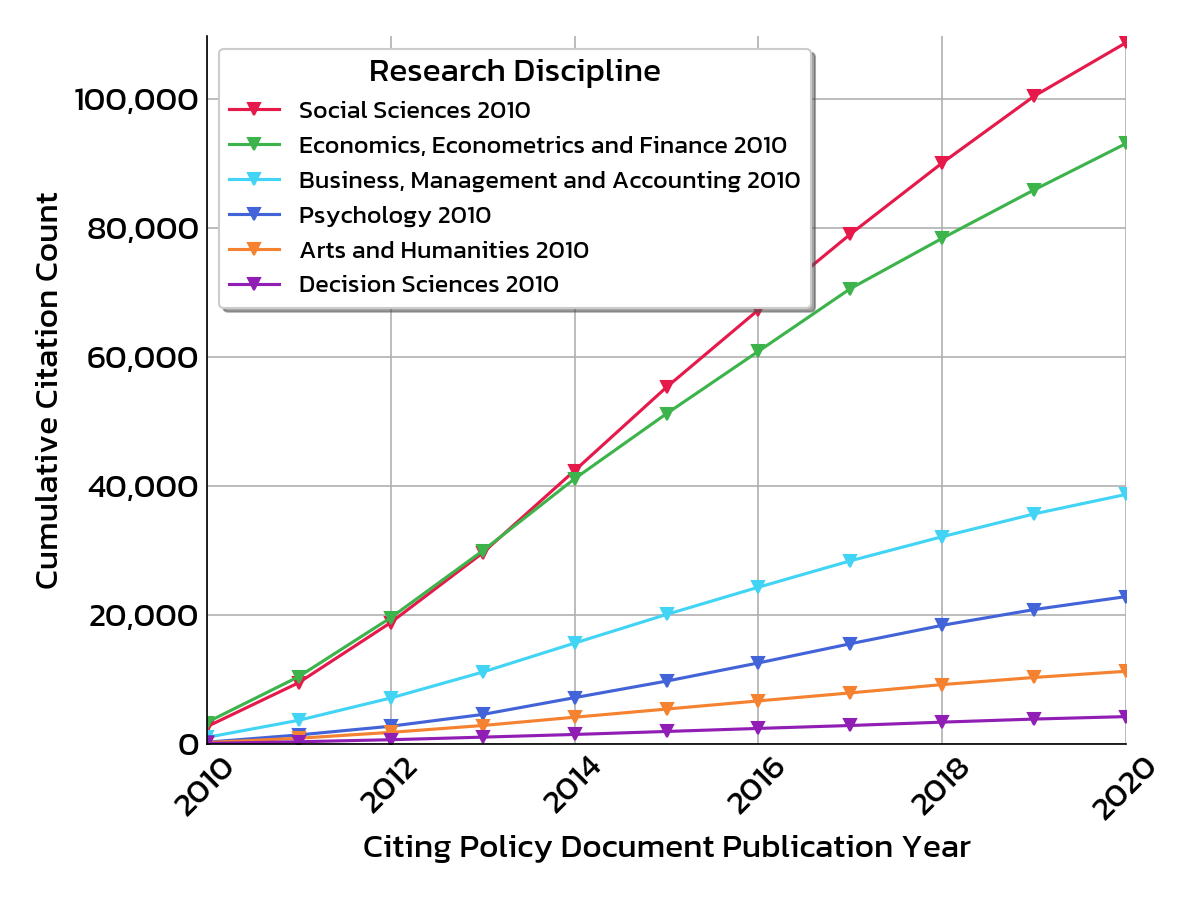}
         \caption{Number of Citations to DOIs published in 2010 in Social Sciences disciplines}
         \label{fig:7c}
     \end{subfigure}
    \caption{Time-series of citations received to DOIs}
    \label{fig:7}     
\end{figure}

Differences in the rate of citation accumulation between different disciplines were also analysed. In terms of broad research areas, Figure \ref{fig:7a} shows cumulative citation rates for articles that were published in 2010. DOIs published in journals categorized as Social Science and Humanities received the most citations, followed by Health Sciences and then Life Sciences. There is marked drop in citation rate for Physical Sciences and Engineering journals. The data for Social Science and Humanities is further decomposed into disciplines in Figure \ref{fig:7c} and reveals most citations in this area are to journals in Social Sciences and Economics fields. This subject balance is in contrast to traditional bibliometric databases which tend to be dominated by citations to papers in biological and physical sciences, but could reasonably be expected given the typical domain of policy setting (e.g. social, economic and environmental).

\subsection{What is the time-lag between the publication of scholarly works and their citation within policy literature and how does this vary between disciplines?}
\begin{figure}[!t]
     \centering
     \begin{subfigure}[b]{.99\textwidth}
         \centering
         \includegraphics[width=\textwidth]{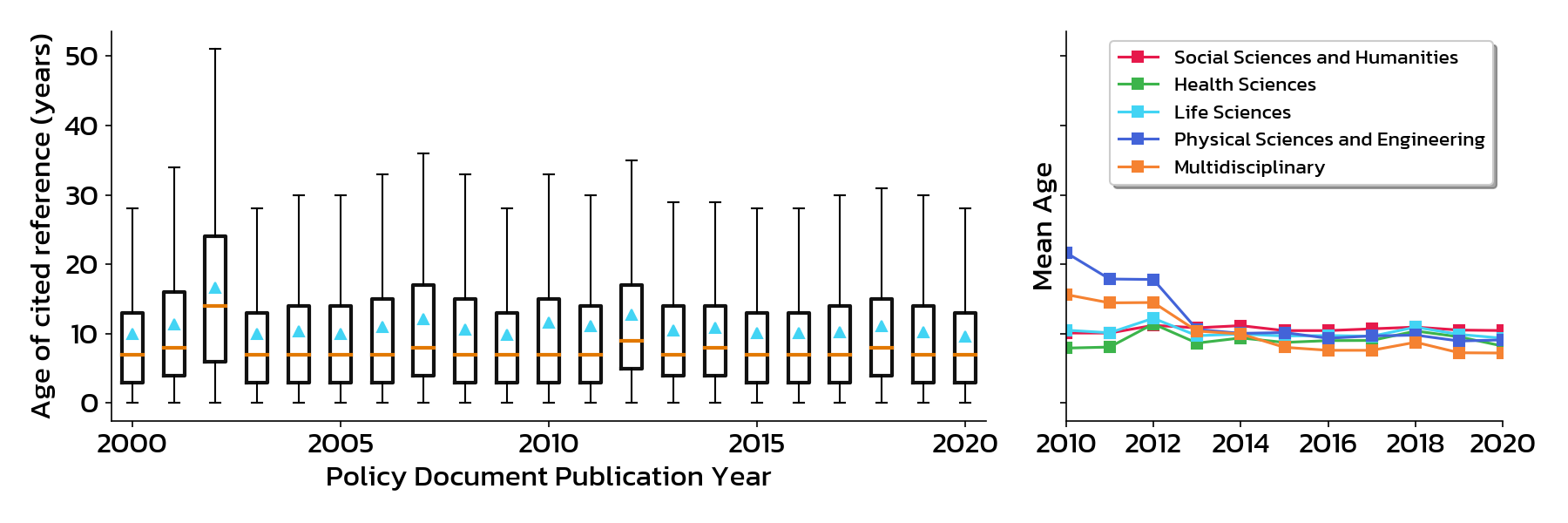}
         \caption{Distribution of cited reference ages aggregated by policy document publication year}
         \label{fig:8a}
     \end{subfigure}
    \vfill     
     \begin{subfigure}[b]{.99\textwidth}
         \centering
         \includegraphics[width=\textwidth]{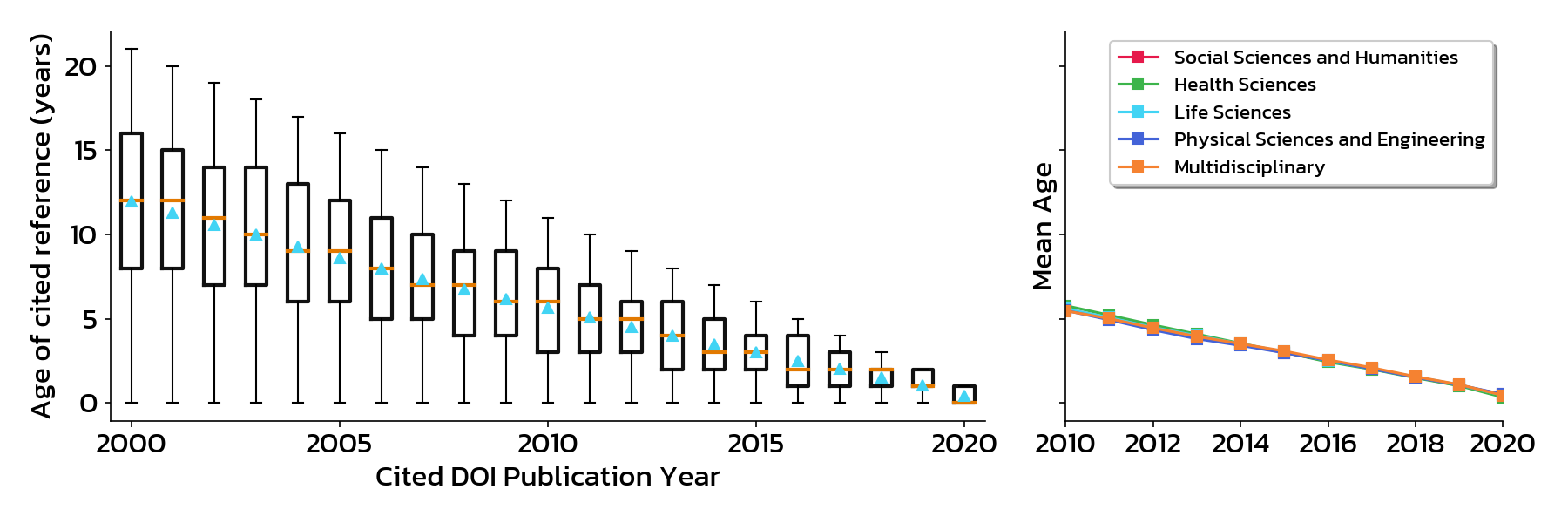}
         \caption{Distribution of cited reference ages aggregated by cited DOI publication year}
         \label{fig:8b}
     \end{subfigure}

    \caption{Age of publications referenced by policy documents}
    \label{fig:8}     
\end{figure}

%figs 8-9
For each year between 2000-20, we analyse the age of cited references in all policy documents indexed. For example, a policy document published in 2015 that references a DOI published in 2010 has a cited reference age of 5 years. For the purposes of this analysis, any reference ages that are calculated to be negative (i.e. the policy document publication date is before that of the cited reference) are removed on the assumption that they represent data errors. The distribution of these ages is displayed using standard box and whisker plots in Figure \ref{fig:8} (orange lines denoting median values, blue triangles for mean). The upper plot (Figure \ref{fig:8a}) aggregates by the publication year of the citing policy document, and the lower plot (Figure \ref{fig:8b}) aggregates by the year of publication for the cited DOI. The right insert in each shows the mean of the distribution for each of the ASJC research areas. Over the 21 year period sampled, there is little variation in the distribution of cited reference ages, with a mean of around 10 years (Figure \ref{fig:8a}), and no significant differences between research areas (right plot). As a result, the distribution of reference ages aggregated by cited DOI publication year (Figure \ref{fig:8b}) shows a consistent trend where the oldest publications have had the longest period to accumulate citations.

\begin{figure}[!t]
     \centering
     \begin{subfigure}[b]{.49\textwidth}
         \centering
         \includegraphics[width=\textwidth]{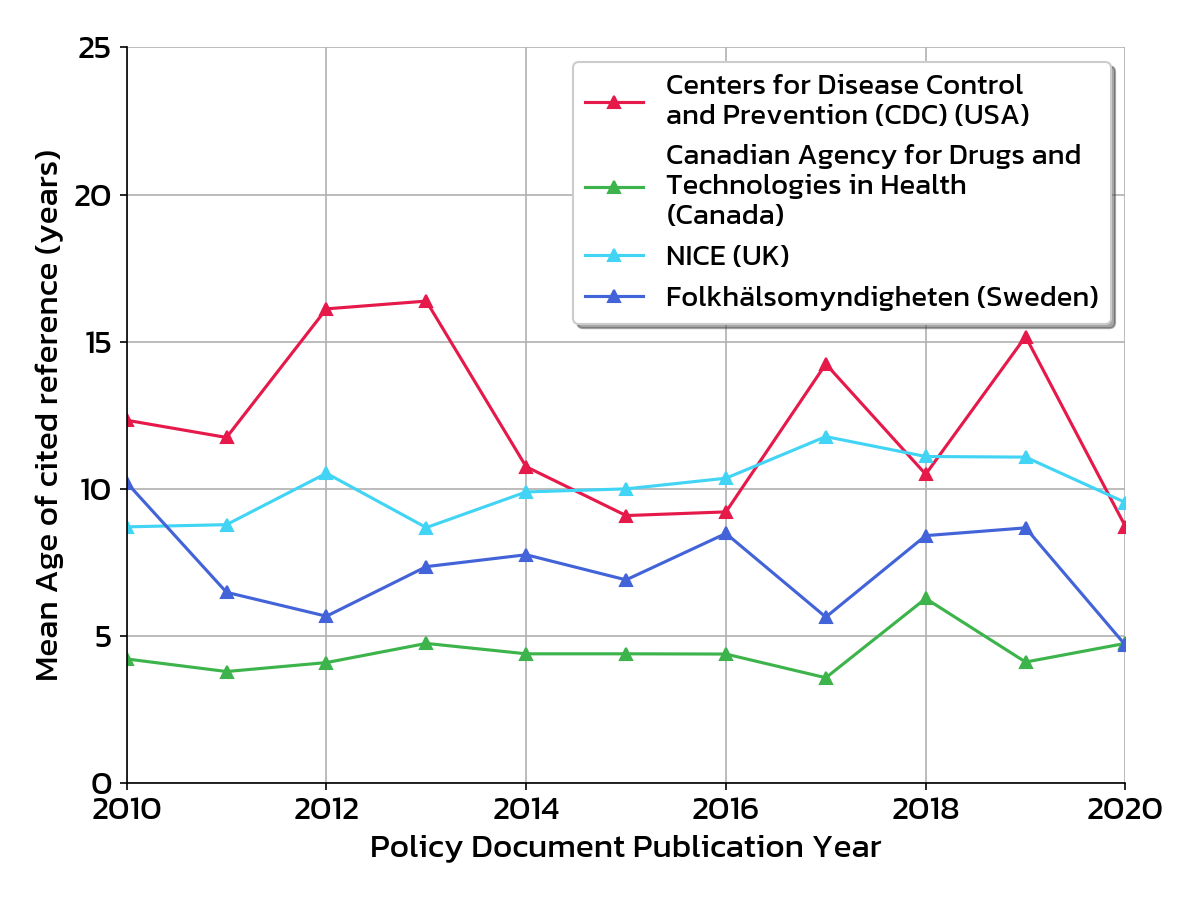}
         \caption{Cited reference age for Healthcare Agencies}
         \label{fig:9a}
     \end{subfigure}
     \hfill     
     \begin{subfigure}[b]{.49\textwidth}
         \centering
         \includegraphics[width=\textwidth]{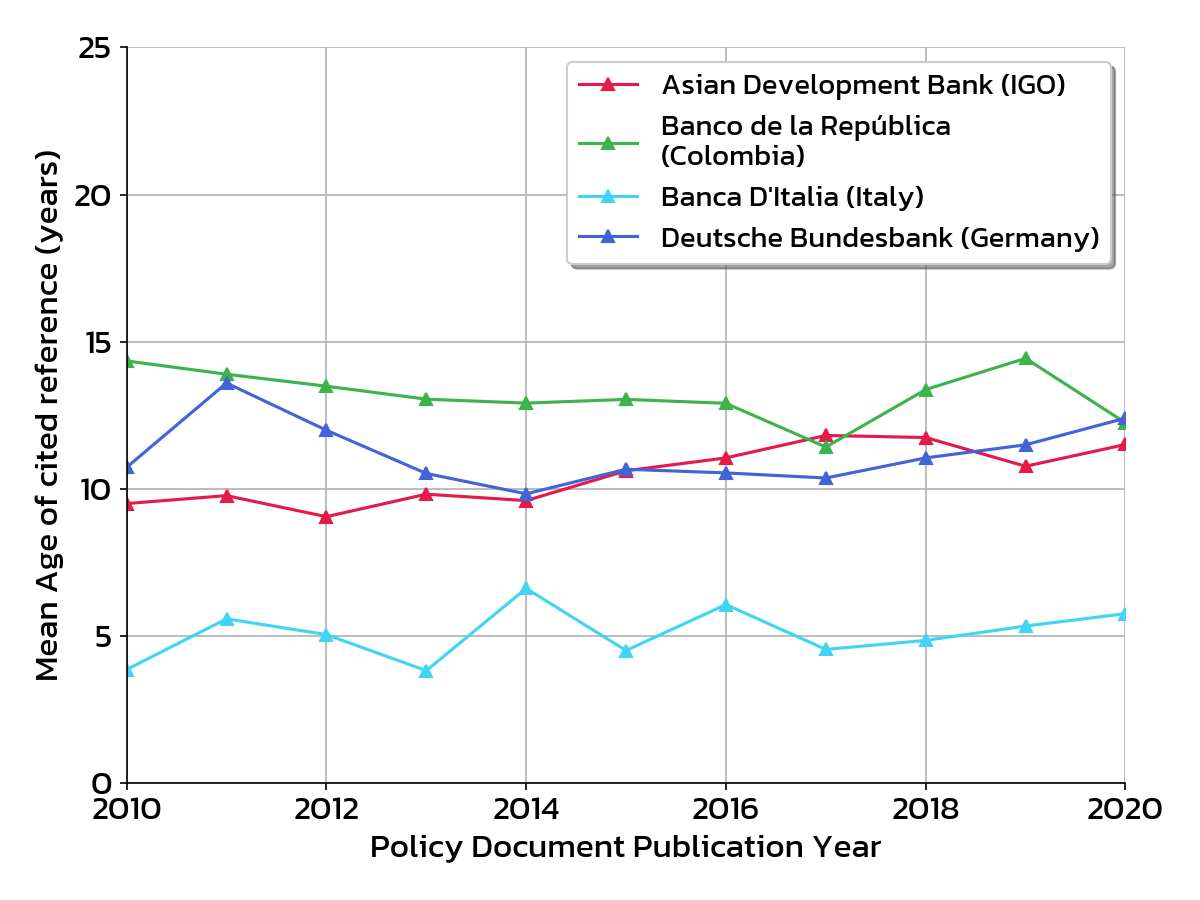}
         \caption{Cited reference age for Government Banks}
         \label{fig:9b}
     \end{subfigure}

    \caption{Age of publications referenced by Policy organization type}
    \label{fig:9}     
\end{figure}

Although cited reference age appears to be consistent at a broad level, we also checked for differences in the age of references between different policy organizations. Two examples are provided in Figure \ref{fig:9} showing four organizations classified as either Healthcare Agency (Figure \ref{fig:9a}) or Government Bank (Figure \ref{fig:9b}). In both of these plots, it is apparent that different organizations cite research with different age ranges. The Canadian Agency for Drugs and Technologies in Health Canada cite much more recent articles on average than the Centers for Disease Control and Prevention (USA). Of course, there are many factors that could influence such a difference, so any interpretation should be mindful of context and comparability of items.

\pagebreak
\subsection{What distribution best models policy citations counts to DOIs?}
%figs 10-11
When examining the policy citation counts of DOIs, it is apparent that the distribution is heavy-tailed \parencite{asmussen_2003}.  For example, for DOIs published between 2010-14 ($n=731,696$), $425,268$ are cited only once (58\%), and only $25,190$ are cited 10 or more times (3.4\%). Prior research using conventional bibliographic databases have investigated possible statistical distributions that model citation data \parencite{eom_2011,brzezinski_2015,thelwall_2016,golosovsky_2021}, although there is some disagreement on whether powerlaw, log-normal or negative binomial distributions are best. Results vary depending on time period and discipline analysed, database used, and if documents with zero citations are included. For this analysis, uncited DOIs are not known since the database is generated by following references made at least once from the policy literature.

\begin{figure}[!t]
\centering
\includegraphics[width=\textwidth]{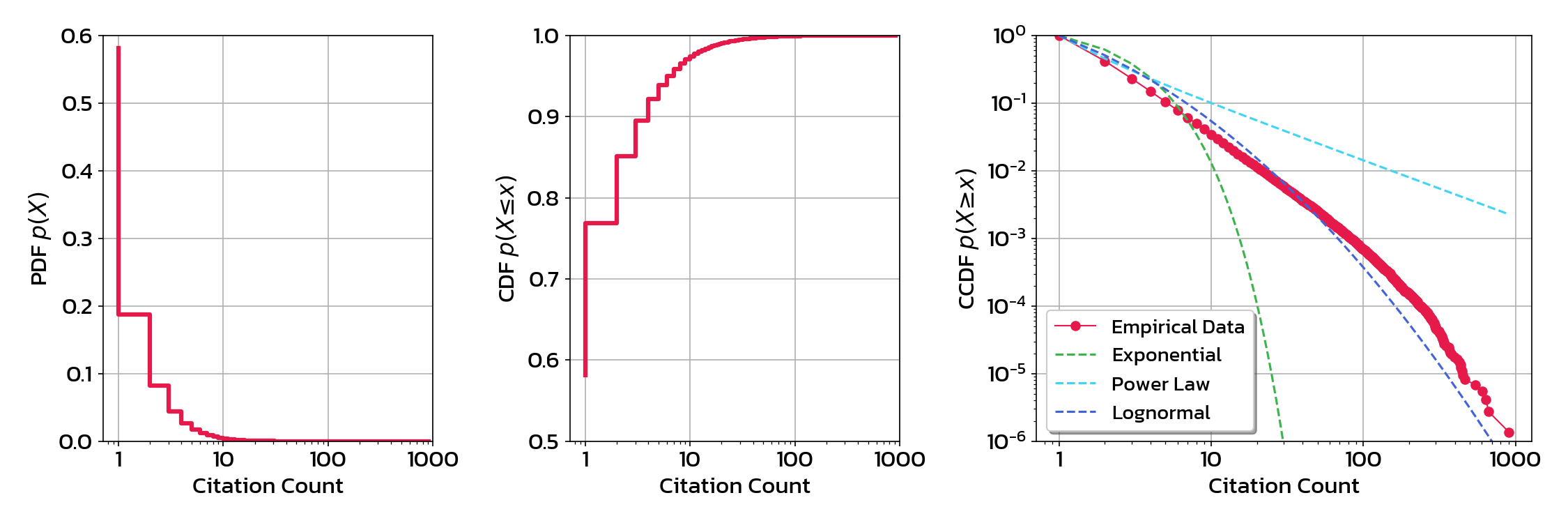}
\caption{Probability distribution functions for DOIs published between 2010-2014}
\label{fig:10}
\end{figure}

Figure \ref{fig:10} provides the probability distribution function (PDF - left), cumulative distribution function (CDF - middle), and complementary cumulative distribution function (CCDF - right) for citations received by DOIs published between 2010-14. We use the Python package Powerlaw \parencite{alstott_2014} to fit distributions to exponential, power law, and lognormal. None of these provide an excellent fit for the data, although lognormal is the closest. In all cases, fitted data overestimate slightly the frequency of low-cited DOIs (i.e. cited fewer than 10 times). Broadly speaking, it appears as though the distribution of policy document citations is similar in nature to that of academic citations.
\begin{figure}[!t]
     \centering
     \begin{subfigure}[b]{.49\textwidth}
         \centering
         \includegraphics[width=\textwidth]{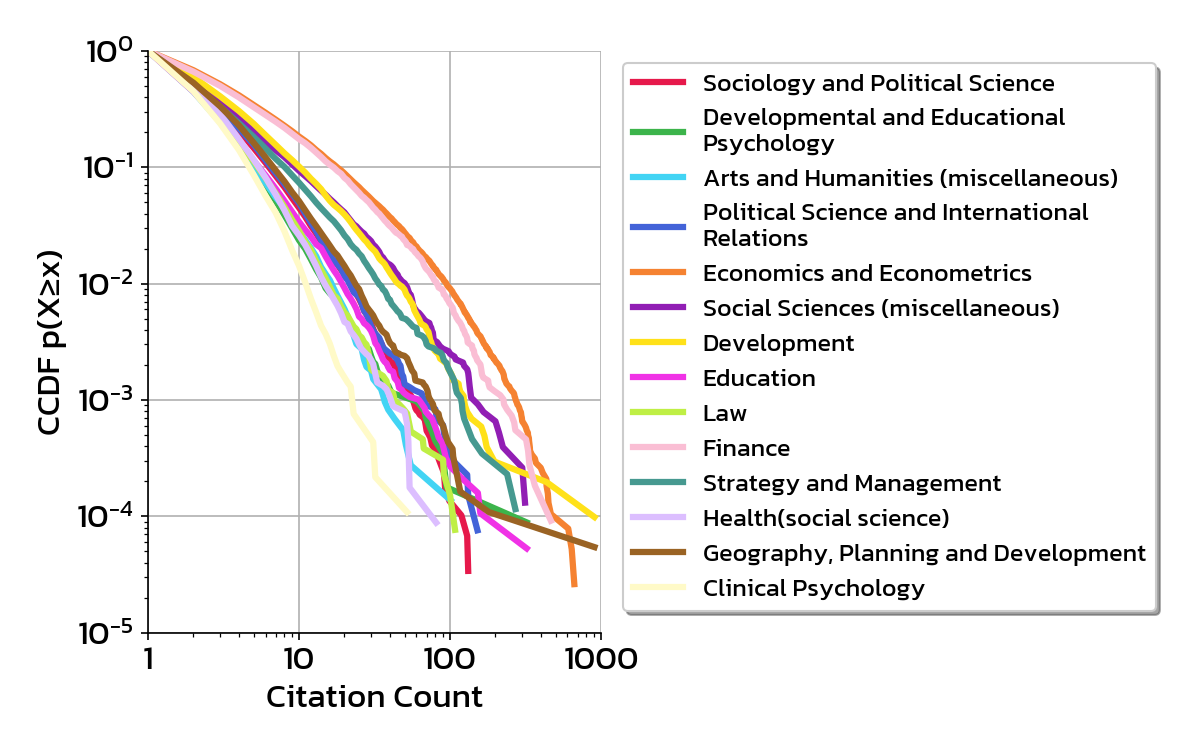}
         \caption{Citations to DOIs published 2010-14 in Social Sciences and Humanities subjects}
         \label{fig:11a}
     \end{subfigure}
     \hfill     
     \begin{subfigure}[b]{.49\textwidth}
         \centering
         \includegraphics[width=\textwidth]{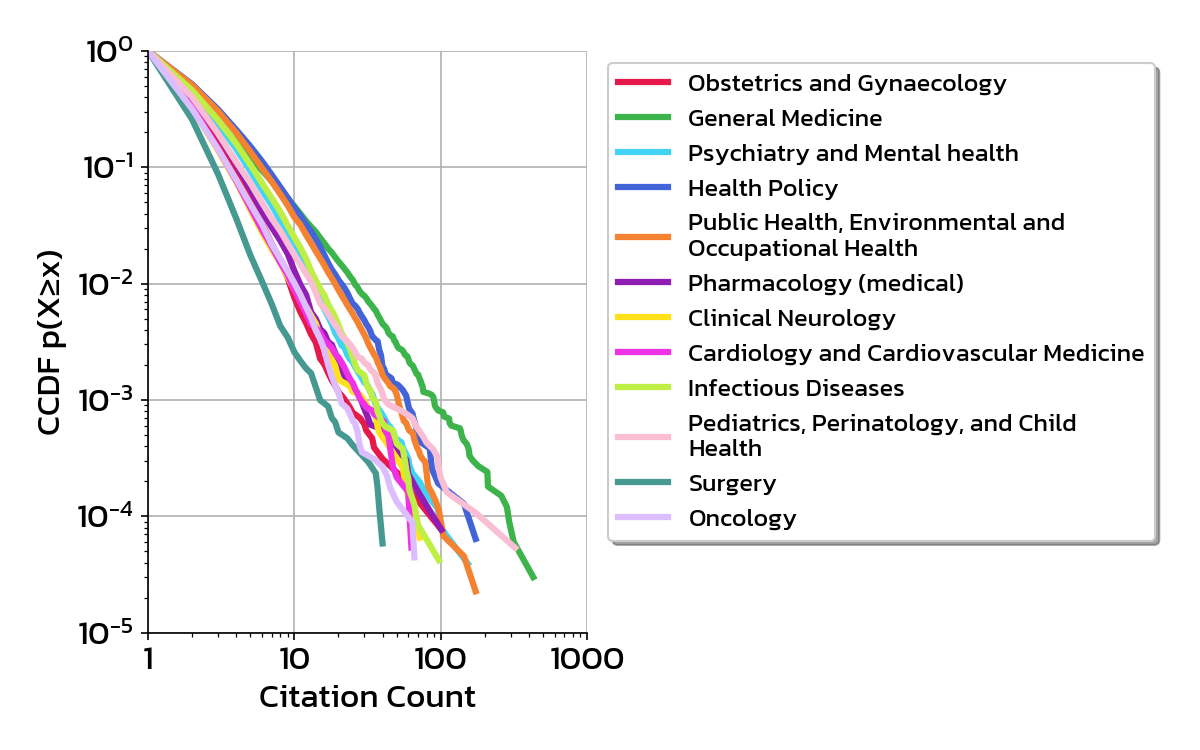}
         \caption{Citations to DOIs published 2010-14 in Health Sciences subjects}
         \label{fig:11b}
     \end{subfigure}

    \caption{Time-series of citations received to DOIs}
    \label{fig:11}     
\end{figure}

Since prior research has shown some variation in citation distributions according to subject \parencite{wallace_2009}, we analysed a sample of subjects from the ASJC research areas `Social Sciences and Humanities' (Figure \ref{fig:11a}) and `Health Sciences' (Figure \ref{fig:11b}). In both cases, it is evident that substantial differences occur between subjects. For example, in the Social Sciences, Economics and Finance receive significantly more citations than in Clinical Psychology or the Arts. This is important to note as it informs the selection of granularity for any field-based normalization. These findings suggest that variation at the subject level is present and therefore, subject-level normalization is preferable, providing sufficiently large benchmark sets can be constructed.

\subsection{How feasible is field-based citation normalization?}
% figs 12 + table 2
As with standard citation metrics, citation counts from policy documents to DOIs also vary according to year of publication and field. Hence, we consider the feasibility of producing field-normalized citation indicators by analysing the number of DOIs cited at least once according to subject and year. From a practical point of view, it is necessary to have a minimum number of DOIs to compare for any combination of subject and publication year. If the data are too sparse (i.e. there are only a handful of DOIs to compare for any subject-year), normalization will not give robust results.

\begin{figure}[!t]
    \centering
    \includegraphics[width=.8\textwidth]{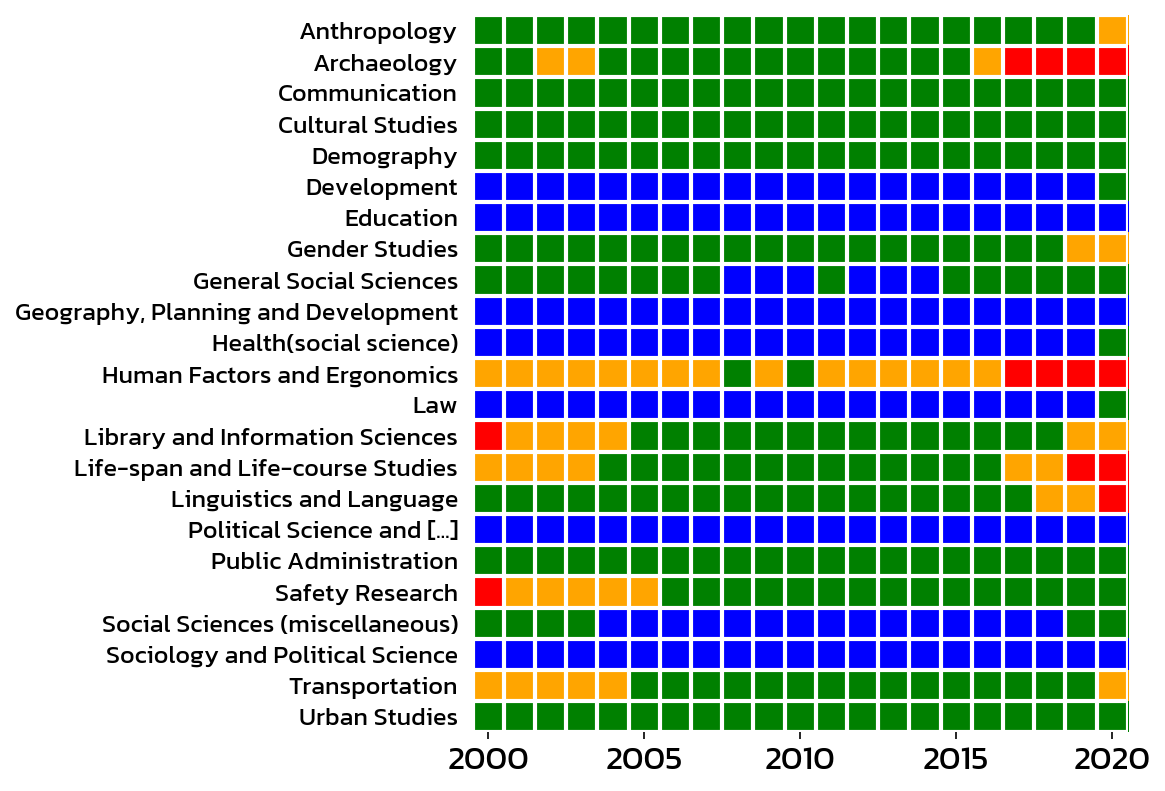}
    \caption{Number of DOIs cited at least once in the discipline `Social Sciences' by subject and year: $s<150$ (red), $150\leq s <250$ (orange), $250\leq s < 1,000$ (green) and $s \geq 1,000$ (blue).}
    \label{fig:12}     
\end{figure}

To illustrate coverage, Figure \ref{fig:12} is provided showing a heatmap of subjects in the discipline `Social Sciences' in terms of the number of DOIs cited each year from 2000-20. The colour coding shows cases where $n$ documents are cited where  $n<150$ (red), $150 \leq n <250$ (orange), $250 \leq n <1,000$ (green), and $n\geq1,000$ (blue). According to \parencite{rogers_2020}, a minimum sample size of 250 is advised for bibliometric samples. The image clearly shows variation in the availability of data. In some subjects, large enough samples could be drawn throughout the study period (e.g. Development. Education, Law), but in other subjects, the data is more sparse and it would be ill-advised to construct normalized indicators (e.g. Human Factors and Ergonomics). As expected, samples sizes are much smaller in the most recent years as these articles are yet to accumulate a significant number of citations.

The above analysis was carried out for all 330 ASJC subjects linked in the data, grouped into 26 disciplines, to determine the overall spread of data availability. For each row in Table \ref{table:2}, a discipline is listed along with:
\begin{itemize}
    \item \textbf{Subjects}: The total number of subjects in the discipline.
    \item \textbf{2000-20\%}: The percentage of subjects where $n \geq 250$ in every year 2000-20.
    \item \textbf{2000-18\%}: The percentage of subjects where $n \geq 250$ in every year 2000-18.
    \item \textbf{years\%}: Across all subjects in the discipline, the percentage of subject-years where  $n \geq 250$.
    \item \textbf{dois\%}: Across all subjects, the percentage of DOIs that are in a subject-year where $n \geq 250$.
\end{itemize}
From this data, it is clear that some disciplines are well covered and others are not. The best covered (i.e. with years\% $> 80$ and dois\% $> 90$) are `Agricultural and Biological Sciences', `Economics, Econometrics and Finance', `Environmental Science', `Immunology and Microbiology', `Medicine', and `Social Sciences'. The least well covered in terms of dois\% are
`Materials Science', `Dentistry', `Physics and Astronomy', `Health Professions', and `Chemical Engineering'. 

Of the $2,270,711$ cited DOIs that were published between 2000-18, $2,009,302$ (88\%) are in a subject that contains at least 250 other cited articles in the same year. This means a subject-level normalisation approach is practical and could be applied to a large portion of scholarly references. 

\begin{table}[!t]
\centering
\begin{tabular}{|l||r|r|r|r|r|}
\hline
                                  Discipline &  Subjects &  2000-20\%         &  2000-18\%          &  years\%           &  dois\%             \\
\hline
        Agricultural and Biological Sciences &            12 &                41.7 &                83.3 &                 88.9 &                 98.3 \\
                         Arts and Humanities &            14 &                14.3 &                14.3 &                 35.7 &                 82.8 \\
Biochemistry, Genetics and Molecular Biology &            16 &                31.2 &                68.8 &                 73.2 &                 95.9 \\
         Business, Management and Accounting &            11 &                54.5 &                63.6 &                 74.5 &                 93.1 \\
                        Chemical Engineering &             9 &                 0.0 &                 0.0 &                 16.4 &                 59.6 \\
                                   Chemistry &             8 &                12.5 &                25.0 &                 41.1 &                 83.1 \\
                            Computer Science &            13 &                 7.7 &                 7.7 &                 27.5 &                 63.8 \\
                           Decision Sciences &             5 &                 0.0 &                20.0 &                 36.2 &                 70.6 \\
                                   Dentistry &             5 &                 0.0 &                 0.0 &                 18.1 &                 52.4 \\
                Earth and Planetary Sciences &            14 &                 7.1 &                42.9 &                 52.0 &                 85.9 \\
         Economics, Econometrics and Finance &             4 &                75.0 &                75.0 &                 95.2 &                 99.6 \\
                                      Energy &             6 &                16.7 &                16.7 &                 53.2 &                 87.1 \\
                                 Engineering &            17 &                 0.0 &                29.4 &                 45.4 &                 82.3 \\
                       Environmental Science &            13 &                84.6 &                92.3 &                 97.8 &                 99.7 \\
                          Health Professions &            16 &                 0.0 &                 6.2 &                  9.2 &                 50.8 \\
                 Immunology and Microbiology &             7 &                57.1 &                71.4 &                 81.6 &                 98.8 \\
                           Materials Science &             9 &                 0.0 &                 0.0 &                 23.3 &                 51.5 \\
                                 Mathematics &            15 &                 0.0 &                 6.7 &                 15.9 &                 62.4 \\
                                    Medicine &            49 &                53.1 &                73.5 &                 82.9 &                 98.8 \\
                           Multidisciplinary &             1 &               100.0 &               100.0 &                100.0 &                100.0 \\
                                Neuroscience &            10 &                10.0 &                30.0 &                 53.3 &                 83.3 \\
                                     Nursing &            23 &                 4.3 &                 8.7 &                 17.2 &                 66.6 \\
  Pharmacology, Toxicology and Pharmaceutics &             6 &                33.3 &                33.3 &                 55.6 &                 94.5 \\
                       Physics and Astronomy &            11 &                 0.0 &                 0.0 &                 19.0 &                 53.4 \\
                                  Psychology &             8 &                62.5 &                62.5 &                 73.8 &                 95.8 \\
                             Social Sciences &            23 &                60.9 &                69.6 &                 87.8 &                 98.1 \\
                                  Veterinary &             5 &                20.0 &                20.0 &                 21.0 &                 74.9 \\
\hline
\end{tabular}
\caption{Completeness of disciplines and their subjects in terms of minimum sample size for normalization}
\label{table:2}
\end{table}

\subsection{Do the citations tracked in the policy literature correlate with policy influence outcomes attributed to funded grants?}
% fig 13 + table 4
To validate the citation data linked via the Overton database, we perform an analysis using data gathered by UK funders from the Gateway to Research (GTR) portal \parencite{ukri_2018}.  Following funding of certain grants in the UK, academics are required to submit feedback using the ResearchFish platform stating publications that resulted from the funding, as well as various research outcomes including engagement activities, intellectual property, spin out companies, clinical trials, and more. One of these categories, policy influence, is used to report various outcomes including citations from policy documents, clinical guidelines, and systematic reviews. Data are collected at the project level, each of which is associated with various DOIs and policy outcomes. For this analysis, a dataset is constructed using all funded grants with a start year between 2014-20, recording the funder and research subjects specified. The funders analysed are: Arts and Humanities Research Council (AHRC), Biotechnology and Biological Sciences Research Council (BBSCR), Engineering and Physical Sciences Research Council (EPSRC), Economic and Social Research Council (ESRC), Medical Research Council (MRC), and Natural Environment Research Council (NERC). 2014 is earliest year surveyed as it is the year that ResearchFish was first adopted across all seven research councils.

For the analysis, data are aggregated at the project level noting the number of DOIs linked to the project, the total number of policy outcomes reported (referred to as \textit{all policy influence}), the number of policy outcomes of the specific type citation (referred to as \textit{citation influence}), and the total number of Overton citations. Effectively, this gives two features to compare - one, self-reported policy outcomes declared by academics, and another by tracking citations from policy documents via the Overton database. If Overton is able to index a sufficiently broad set of materials, these two features should be correlated. 

Table \ref{table:4} provides the correlation statistics (as measured using Pearson) for the complete dataset (All row), and for each research council. In every row, the total number of projects and DOIs they link to  is reported (columns Projects and DOIs), along with two sets of statistics - one testing Overton citation counts against the total number of policy influence outcomes reported (All policy influence - middle columns), and the other testing Overton citation counts against the number of policy influence outcomes that are specifically for citations in policy documents, clinical guidelines, or systematic reviews (Citation influence only - right columns). Both the correlation coefficient $r$ and pvalue are listed, as well as the percentage of projects that were linked to any policy influence outcomes. In all cases, pvalues are very small (e.g. $10\textsuperscript{-30}$) and are only listed to two decimal places. This percentage figure is given to contextualize results as for some funders, the number of projects associated with any policy outcomes is low. According to these results, the correlation between the count of policy influence outcomes and the total number of citations in Overton is larger when considering all policy influence types, rather than only those specifically for citation, although for EPSRC they are similar, and for ESRC they are higher ($r=0.70$). There is a medium correlation over all funders ($r=0.42$), and good correlation for the EPSRC ($r=0.66$), ESRC ($r=0.48$), and MRC ($r=0.63$).
{\setlength{\tabcolsep}{0.3em}
\begin{table}[!t]
\centering
\begin{tabular}{|lrr||r|rr||r|rr|}
\hline
     &       &        & \multicolumn{3}{l||}{All policy influence} & \multicolumn{3}{l|}{Citation influence only} \\
Funder & Projects & DOIs & r &  pval & Projects\% & r & pval & Projects\% \\
\hline
All & 67,702 & 383,642 &     0.42 &   0.00 &             7.13 &     0.32 &   0.00 &             1.17 \\
AHRC & 3,902  & 14,254  &     0.26 &   0.00 &            13.84 &     0.19 &   0.00 &             2.26 \\
BBSRC & 9,031  & 40,642  &     0.30 &   0.00 &             7.60 &     0.23 &   0.00 &             0.68 \\
EPSRC & 17,799 & 106,312 &     0.66 &   0.00 &             4.72 &     0.65 &   0.00 &             0.51 \\
ESRC & 5,732  & 37,503  &     0.48 &   0.00 &            16.99 &     0.70 &   0.00 &             4.41 \\
MRC & 5,992  & 60,854  &     0.63 &   0.00 &            16.41 &     0.20 &   0.00 &             2.42 \\
NERC & 4,727  & 30,035  &     0.22 &   0.00 &            13.71 &     0.17 &   0.00 &             3.13 \\
\hline
\end{tabular}
\caption{Pearson correlation between funded projects with policy influence and total policy citations in Overton}
\label{table:4}
\end{table}}

\begin{figure}[!p]
    \centering
    \includegraphics[width=\textwidth]{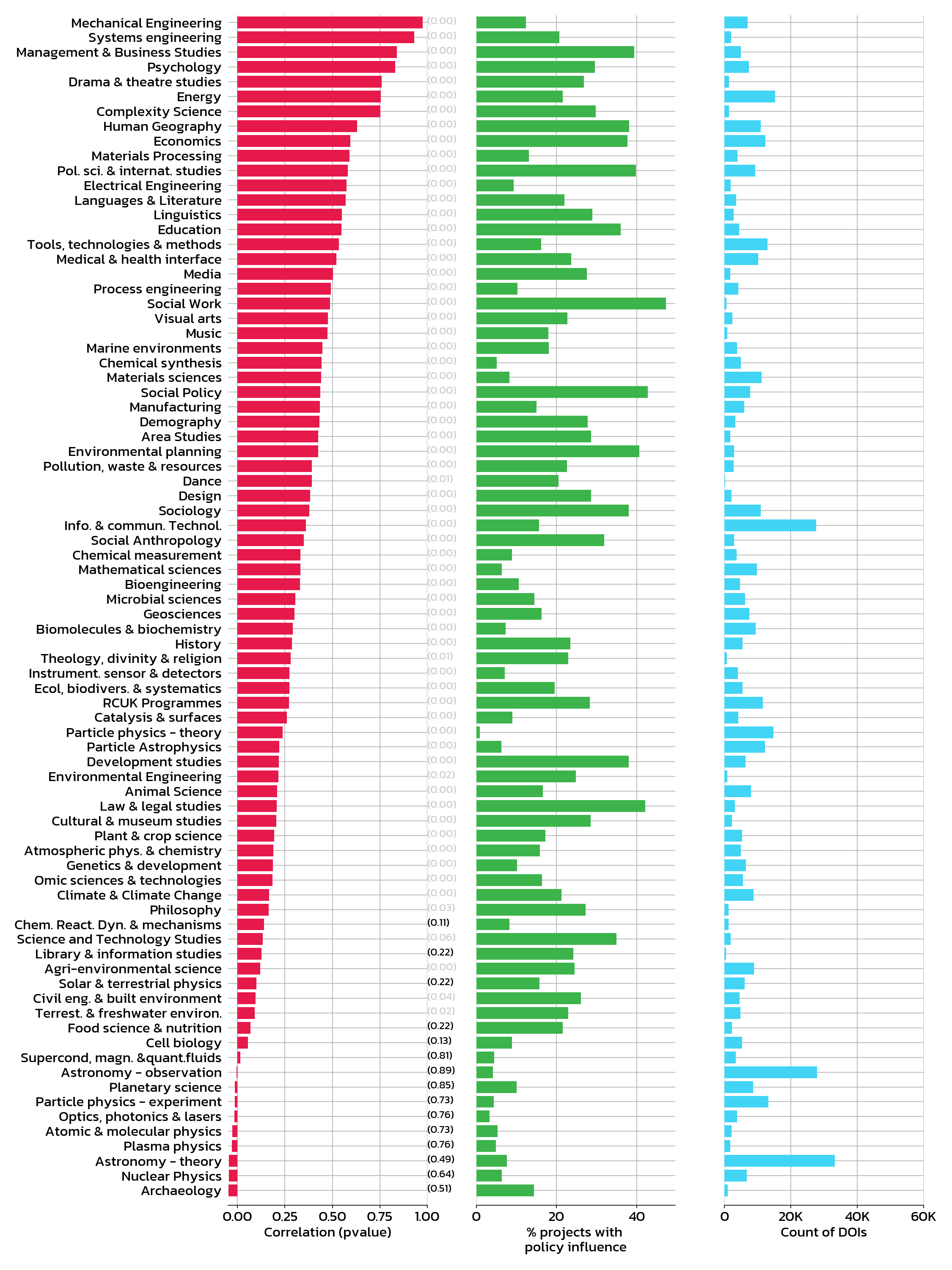}
    \caption{Number of Citations to DOIs by research area}
    \label{fig:13}     
\end{figure}

The data are further decomposed according to subject category assigned to the grant, as depicted in Figure \ref{fig:13}. Each grant may be assigned to multiple subjects and is considered in the calculation for each subject. For each subject (a row), three columns are used to show the correlation $r$ (red), percentage of projects reporting any policy influence (green), and the total count of DOIs linked to projects (blue). Pvalues for correlation statistics are listed in parenthesis and highlighted when $p > 0.05$ (i.e. when the the correlation is \textbf{not} statistically significant). In this plot, correlations are measured against all policy influence outcomes (i.e. corresponding to the middle columns in Table \ref{table:4}). When analysed at this level of granularity, there is a large spread in the Pearson correlation, although all statistically significant correlations are positive. 17 subjects have a correlation $> 0.5$, but 39 have a correlation $< 0.3$. 

These results show that for some subjects, Overton citation data correlates well with policy influence outcomes reported by academics. This occurs most in subjects that might be expected to have some policy influence, such as Management \& Business Studies ($r=0.84$), Psychology ($r=0.83$), Human Geography ($r=0.63$), Economics ($r=0.60$), and Political Science \& International Studies ($r=0.58$), but also in others that might not, such as Mechanical Engineering ($r=0.98$), Systems engineering ($r=0.93$), and Drama \& Theatre Studies ($r=0.76$).

\pagebreak
\subsection{Does the amount of policy citation correlate with peer-review assessment scores as reported in the UK REF2014 impact case study data?}
% table 5 6
To test for possible correlation, we utilize the Impact Case Study database from REF2014. This contains $6,637$ 4-page documents that outline the wider socio-economic and cultural impact of research attributed to a particular university and UoA. Part of the case study document references the original underpinning research (up to six references per case study) which has been linked via DOIs. By means of peer-review, each case study is scored as 4* (world-leading), 3* (internationally excellent), 2* (internationally recognised), or 1* (nationally recognised). Although the scores for individual cases studies are not known, the aggregate scores are made available as the percentage of case studies that received each score. Hence, it is possible to test possible correlations at the aggregate level (namely, institution and UoA).

For this analysis, we test the correlation between research scored as 4* (excellent) and citations to the underpinning research as reported in the Overton database. Since the assessment exercise took place in 2014, only citations from policy documents published in or earlier than 2014 are considered. Rather than test raw citation counts, we calculate a year-subject normalized citation percentile for each DOI using ASJC journal categories (i.e. all DOIs published in a certain year and subject are compared with each other). Any DOIs in a year-subject group that contain $<250$ examples are marked as invalid and excluded from the analysis. Of the total $24,945$ unique DOIs associated with an impact case study, $4,292$ are referenced in Overton and have a valid citation percentile.

Following the methodology presented in \parencite{traag_2019}, we measure the correlation between the percentage of case studies that scored 4* and the percentage of DOIs in the top 99, 90, and 75\textsuperscript{th} Overton citation percentiles. Multiple percentiles were tested as it it not necessarily clear where the benchmark for 4* research would lie. A total of $1,847$ scores are evaluated - one for each university and UoA. A size-independent test measures the Pearson correlation between the percentage of research scored 4* and the percentage of DOIs with a normalized citation percentile above the threshold. 

Table \ref{table:5} provides the results of this analysis. All 36 units of assessment are shown along with the Pearson correlation $r$ and pvalue for three citation percentile thresholds: 99\%, 90\%, and 75\%. In some cases (for example Classics), when no DOIs could be found exceeding the percentile threshold, the correlation is undefined and hence, left blank. Based on these results, it is apparent that different percentile thresholds yield different results depending on UoA. For example in UoAs 18 Economics and Econometrics and 25 - Education, the highest statistically significant correlations of 0.52 and 0.46 respectively are obtained with a threshold of 90\%, but in UoA 7 - Earth Systems and Environmental Sciences, a threshold of 99\% yields the highest correlation of 0.52. This suggests that the threshold for what is considered 4* impact varies across fields in terms of policy influence.

This analysis shows that for some UoAs, Overton policy citation percentiles do correlate with peer-review assessment, but less than reported for citation data \parencite{traag_2019} when compared to scoring of outputs. For many UoAs, no correlation can be inferred due to large pvalues. Ideally, the test would only be performed on the subset of case studies that might reasonably be expected to have some form of policy outcome. For example, searching the database for \texttt{"policy outcome"$\sim$5 OR "policy influence"$\sim$5} (where the $\sim$5 operator specifies that terms must be within five words of each other) only returns 406 results. Hence, our test effectively measures the correlation of impact in general against that of policy citation and could only be expected to find correlation in UoAs where the dominant form of impact is policy related, such as in UoA 22 Social Work and Social Policy. Unfortunately, because scores are not known for individual case studies, this type of analysis is not possible.

{\setlength{\tabcolsep}{0.3em}
\begin{table}[!h]
\centering
\begin{tabular}{|l||r|r||r|r||r|r|}
\hline
{} & \multicolumn{2}{l||}{4*-top99} & \multicolumn{2}{l||}{4*-top90} & \multicolumn{2}{l|}{4*-top75} \\
uoa & r & pval & r & pval & r & pval \\
\hline
1 - Clinical Medicine                   &     0.20 & 0.29 &     0.24 & 0.20 &     0.25 & 0.17 \\
2 - Public Health, Health [...]         &     0.22 & 0.23 &    -0.20 & 0.27 &     0.24 & 0.18 \\
3 - Allied Health Professions, [...]    &     0.02 & 0.84 &     0.04 & 0.73 &     0.13 & 0.23 \\
4 - Psychology, Psychiatry and [...]    &     0.18 & 0.11 &     0.13 & 0.25 &     \color{blue}0.27 & \color{blue}0.01 \\
5 - Biological Sciences                 &     0.16 & 0.30 &     0.08 & 0.59 &     0.02 & 0.89 \\
6 - Agriculture, Veterinary and [...]   &     0.28 & 0.16 &     \color{blue}0.57 & \color{blue}0.00 &     \color{blue}0.54 & \color{blue}0.00 \\
7 - Earth Systems and [...]             &     \color{blue}0.52 & \color{blue}0.00 &     0.24 & 0.11 &     0.17 & 0.26 \\
8 - Chemistry                           &          &      &     0.15 & 0.39 &     0.00 & 0.99 \\
9 - Physics                             &     0.07 & 0.69 &     0.07 & 0.69 &     0.02 & 0.91 \\
10 - Mathematical Sciences              &     \color{blue}0.32 & \color{blue}0.02 &     0.11 & 0.44 &     0.13 & 0.36 \\
11 - Computer Science and Informatics   &          &      &     \color{blue}0.27 & \color{blue}0.01 &     \color{blue}0.30 & \color{blue}0.01 \\
12 - Aeronautical, Mechanical, [...]    &          &      &     \color{blue}0.49 & \color{blue}0.02 &     \color{blue}0.64 & \color{blue}0.00 \\
13 - Electrical and Electronic [...]    &          &      &     \color{blue}0.42 & \color{blue}0.01 &     0.11 & 0.50 \\
14 - Civil and Construction Engineering &     0.17 & 0.58 &    -0.22 & 0.46 &    -0.11 & 0.72 \\
15 - General Engineering                &          &      &     0.11 & 0.41 &     0.15 & 0.24 \\
16 - Architecture, Built [...]          &     0.13 & 0.40 &     \color{blue}0.34 & \color{blue}0.03 &     \color{blue}0.30 & \color{blue}0.05 \\
17 - Geography, Environmental [...]     &     0.21 & 0.07 &     0.16 & 0.18 &     0.20 & 0.08 \\
18 - Economics and Econometrics         &     \color{blue}0.39 & \color{blue}0.04 &     \color{blue}0.52 & \color{blue}0.00 &     0.33 & 0.09 \\
19 - Business and Management Studies    &    -0.00 & 0.96 &     0.10 & 0.35 &     0.18 & 0.07 \\
20 - Law                                &          &      &     0.16 & 0.19 &     0.05 & 0.69 \\
21 - Politics and International Studies &     0.16 & 0.23 &     0.07 & 0.60 &     0.26 & 0.06 \\
22 - Social Work and Social Policy      &     \color{blue}0.47 & \color{blue}0.00 &     0.24 & 0.06 &     \color{blue}0.32 & \color{blue}0.01 \\
23 - Sociology                          &     0.01 & 0.95 &     0.07 & 0.73 &     0.08 & 0.68 \\
24 - Anthropology and Development [...] &     0.10 & 0.62 &     0.15 & 0.48 &     0.21 & 0.30 \\
25 - Education                          &     \color{blue}0.33 & \color{blue}0.00 &     \color{blue}0.46 & \color{blue}0.00 &     \color{blue}0.40 & \color{blue}0.00 \\
26 - Sport and Exercise Sciences, [...] &     0.22 & 0.12 &     \color{blue}0.40 & \color{blue}0.00 &     \color{blue}0.31 & \color{blue}0.03 \\
27 - Area Studies                       &          &      &     \color{blue}0.44 & \color{blue}0.04 &     \color{blue}0.42 & \color{blue}0.05 \\
28 - Modern Languages and Linguistics   &     0.12 & 0.39 &     0.24 & 0.08 &     0.24 & 0.08 \\
29 - English Language and Literature    &          &      &    -0.00 & 0.99 &    -0.00 & 0.99 \\
30 - History                            &          &      &     0.12 & 0.30 &     0.13 & 0.25 \\
31 - Classics                           &          &      &          &      &          &      \\
32 - Philosophy                         &          &      &     0.24 & 0.13 &     0.20 & 0.21 \\
33 - Theology and Religious Studies     &          &      &          &      &     \color{blue}0.41 & \color{blue}0.02 \\
34 - Art and Design: History, [...]     &          &      &          &      &    -0.15 & 0.22 \\
35 - Music, Drama, Dance and [...]      &          &      &    -0.01 & 0.97 &     0.20 & 0.09 \\
36 - Communication, Cultural and [...]  &          &      &     \color{blue}0.38 & \color{blue}0.00 &     \color{blue}0.33 & \color{blue}0.01 \\
\hline
\end{tabular}
\caption{Pearson correlation for REF2014 impact scores marked 4* against Overton citation percentiles above three threshold values: 99\%, 90\% and 75\%. Correlations with a pvalue $<0.05$ are highlighted in blue.} 
\label{table:5}
\end{table}}

\pagebreak
\section{Discussion}  \label{sec:discussion}
Our analysis of the Overton policy document citation database yields a promising outlook.  Using this kind of data, it is possible to link the original research published in scholarly literature to their use in a policy setting environment. The Overton database indexes a sufficient amount of content to create large volumes of citations ($>400,000$ every year since 2014) across a wide range of research topics and journals. Unlike conventional bibliometric databases, citations are more focused towards social sciences, economics and environmental sciences than to biological and physical sciences, a feature that suggests novel value in the content in terms of analytical potential.

The balance of content by region broadly follows that of other bibliometric databases, namely it is dominated by North America and Europe, but the representation of local language documents is much higher than in scholarly publishing where English dominates \parencite{mongeon_2016,marquez_2020}. Anecdotal evidence in this study hints that Overton may have more equitable coverage across some countries: Figure \ref{fig:3} shows that Peru and Uruguay have a similar volume of policy documents indexed to Brazil and Chile despite producing fewer scholarly works. However, more detailed analysis drawing on other indicators (e.g. economic and industrial) is required to produce robust conclusions in relation to this question.

Although a significant proportion of the policy documents indexed are not linked to DOIs (88\% of PDFs), a core set of around $200,000$ contain more than 8M references. This reflects the diverse range of material indexed including statistical reports, informal communications, proceedings, and commentary, many of which one would not expect to contain references to original research articles. A considerable pool of citations is generated - between $200,000$ and $400,000$ per year since 2000 across a broad set of journals. A more detailed analysis of this data could compare how citations are distributed across journals and if citations patterns from policy documents follow the same tendencies as scholarly publishing. It may be true that some journals are able to demonstrate higher utilisation in policy documents relative to a citation-based ranking.

The potential for development of field-normalized citation indicators is good. When analysed at the ASJC subject level, many fields contain a sufficient number of cited articles to create benchmarks (i.e. $\geq250$), especially if the most recent two years are excluded. Overall, 88\% of articles published between 2000-18 that receive any policy citations could be field-normalized in this way. However, although this approach is practical, it may not be best - a more detailed analysis comparing normalization results at different levels of granularity (i.e. field-based or discipline-based) would be required to make any recommendation.

One potentially interesting line of enquiry is that of citation lag. At the macro scale, our analysis shows there is little variation in the distribution of ages, even across disciplines, but when viewed at a more granular level (such as individual policy organizations), diversity occurs. This may offer useful insights into the differences between what research is used, in terms of age and also in citation ranking. Some organization may favour newer but less established evidence than others that prefer older but more widely recognized research.

The distribution of citations accumulated by research articles seems to follow similar trends to that seen in other altmetric indicators, especially Mendeley, Twitter and Facebook as reported in \parencite{fang_2020}, and like conventional citation data, are best matched to a log-normal distribution. It is interesting to note that in \parencite{fang_2020}, $12,271,991$ articles published between 2012 and 2018 were matched to Altmetric data and yielded $156,813$ citations across $137,326$ unique documents. For the same time period, Overton contains $2,600,464$ citations across $1,006,439$ unique DOIs. These coverage statistics are not directly comparable because the original pool of articles surveyed in \parencite{fang_2020} is limited to the Web of Science and Overton tracks citations to any journal. Nevertheless, it does suggest that Overton tracks substantially more citations to policy literature than Altmetric.

Possibly the most striking and encouraging result is from the analysis of policy influence outcomes reported to UK Funders. Our findings show that for some subjects, correlation between self-reported data and that extracted from Overton is high. This offers additional opportunities to reduce reporting burden, either through semi-automated or automated approaches. Further, it provides a basis to benchmark funders and institutions from different regions where self-reported data may not be available, although such an analysis should consider coverage variation across geographies. 

Finally, the analysis between peer-review assessment and policy citation impact hints at some utility: for certain Units of Assessment, a correlation between peer-review score of impact and citation rank does exist, although less than that seen in other studies that assessed peer-review scores of academic impact against conventional citation data \parencite{traag_2019}. While the REF2014 impact case study data does provide a unique opportunity to understand how research is assessed from the perspective of wider socio-economic impact, obfuscation of the individual scores prevents deeper analysis that is focused on research pertinent to policy outcomes. It may be more fruitful to utilise other sources to benchmark peer-review, such as post-publication peer review score \parencite{waltman_2014}. 

\section{Acknowledgment} 
This research has been funded by Open Policy Ltd who run Overton.

\printbibliography
\end{document}